\documentclass[twocolumn]{aastex701}

\usepackage{lipsum}
\usepackage{comment}
\usepackage{amsmath}
\usepackage{graphicx}
\usepackage{xcolor}
\usepackage{makecell}

\begin{document}

\title{Extreme Enrichment at Cosmic Dawn: Three FeII-Strong Quasars at $z > 6$ from the JWST Aether survey}

\author[0009-0009-8274-441X]{Anniek J. Gloudemans}
\affiliation{NSF NOIRLab, Gemini Observatory, 
670 N A'ohoku Place
Hilo, HI 96720, USA; anniek.gloudemans@noirlab.edu}
\email{anniek.gloudemans@noirlab.edu}  

\author[0000-0002-6822-2254]{Emanuele Paolo Farina}
\affiliation{International Gemini Observatory/NSF NOIRLab, 670 N A’ohoku Place, Hilo, Hawai'i 96720, USA}
\affiliation{INAF – Osservatorio di Astrofisica e Scienza dello Spazio di Bologna, via Gobetti 93/3, I-40129, Bologna, Italy} 
\email{} 

\author[0000-0002-2662-8803]{Roberto Decarli}
\affiliation{INAF – Osservatorio di Astrofisica e Scienza dello Spazio di Bologna, via Gobetti 93/3, I-40129, Bologna, Italy}  
\email{} 

\author[0009-0008-2205-7725]{Klaudia Protu\v{s}ov\'{a}}
\affiliation{Institute for Theoretical Physics, Heidelberg University, Philosophenweg 12, 69120, Heidelberg, Germany}  
\email{} 

\author[0000-0002-4770-6137]{Fabrizio Arrigoni Battaia}
\affiliation{Max-Planck-Institut für Astrophysik, Karl-Schwarzschild-Str. 1, D-85748 Garching bei München, Germany}  
\email{} 

\author[0000-0002-2931-7824]{Eduardo Ba\~{n}ados}
\affiliation{Max-Planck-Institut f\"ur Astronomie, K\"onigstuhl 17, D-69117, Heidelberg, Germany}
\email{} 

\author[0000-0002-3026-0562]{Aaron J.\ Barth}
\affiliation{Department of Physics and Astronomy, 4129 Frederick Reines Hall, University of California, Irvine, CA, 92697-4575, USA}
\email{} 

\author[0000-0003-4747-4484]{Silvia Belladitta}
\affiliation{Max-Planck-Institut f\"ur Astronomie, K\"onigstuhl 17, D-69117, Heidelberg, Germany}
\affiliation{INAF-Osservatorio di Astrofisica e Scienza dello Spazio, via Gobetti 93/3, I-40129, Bologna, Italy}
\email{} 

\author[0000-0002-4314-021X]{Manuela Bischetti}
\affiliation{Dipartimento di Fisica Enrico Fermi, Università di Pisa, Largo Bruno Pontecorvo 3, Pisa, I-56127, Italy}
\affiliation{INAF -- Osservatorio Astronomico di Trieste, Via G. B. Tiepolo 11, I-34131 Trieste, Italy}
\email{} 

\author[0000-0002-3173-1098]{Hyunseop Choi}
\affiliation{Department of Astronomy, University of Michigan, 1085 S. University Ave., Ann Arbor, MI 48109, USA}
\email{} 

\author[0000-0001-8986-5235]{Dominika \v{D}urov\v{c}\'{i}kov\'{a}}
\affiliation{MIT Kavli Institute for Astrophysics and Space Research, Massachusetts Institute of Technology, Cambridge, MA 02139, USA}  
\affiliation{Department of Physics, Massachusetts Institute of Technology, Cambridge, MA 02139, USA}  
\email{} 

\author[0000-0003-2895-6218]{Anna-Christina Eilers}
\affiliation{MIT Kavli Institute for Astrophysics and Space Research, Massachusetts Institute of Technology, Cambridge, MA 02139, USA}  
\affiliation{Department of Physics, Massachusetts Institute of Technology, Cambridge, MA 02139, USA}  
\email{} 

\author[0000-0002-7200-8293]{Simona Gallerani}
\affiliation{Scuola Normale Superiore, Piazza dei Cavalieri 7, 56126 Pisa, Italy}  
\email{} 

\author[0000-0001-6179-7701]{Thales A.\ Gutcke}
\affiliation{Institute for Astronomy, University of Hawai`i, 2680 Woodlawn Drive, Honolulu, HI 96822, USA}
\email{} 

\author[0000-0003-1516-9450]{Luca Ighina}
\affiliation{Center for Astrophysics — Harvard \& Smithsonian, 60 Garden St., Cambridge, MA 02138, USA}  
\affiliation{INAF, Osservatorio Astronomico di Brera, via Brera 28, 20121, Milano, Italy}
\email{} 

\author[0000-0002-1428-7036]{Brian C.\ Lemaux}
\affiliation{International Gemini Observatory/NSF NOIRLab, 670 N A’ohoku Place, Hilo, Hawai'i 96720, USA}
\affiliation{Department of Physics and Astronomy, University of California, Davis, One Shields Ave., Davis, CA 95616, USA}
\email{} 

\author[0000-0002-8858-6784]{Federica Loiacono}
\affiliation{INAF – Osservatorio di Astrofisica e Scienza dello Spazio, Via Gobetti 93/3, I-40129, Bologna, Italy}  
\email{} 

\author[0000-0001-5063-0340]{Yoshiki Matsuoka}
\affiliation{Research Center for Space and Cosmic Evolution, Ehime University, Matsuyama, Ehime 790-8577, Japan}  
\email{} 

\author[0000-0002-5941-5214]{Chiara Mazzucchelli}
\affiliation{Instituto de Estudios Astrof\'{\i}sicos, Facultad de Ingenier\'{\i}a y Ciencias, Universidad Diego Portales, Avenida Ej\'{e}rcito Libertador 441, Santiago, Chile}
\email{} 

\author[0009-0009-1715-4157]{Silvia Onorato}
\affiliation{International Gemini Observatory/NSF NOIRLab, 670 N A’ohoku Place, Hilo, Hawai'i 96720, USA}
\email{} 

\author[0000-0002-2536-1633]{Hyewon Suh}
\affiliation{International Gemini Observatory/NSF NOIRLab, 670 N A’ohoku Place, Hilo, Hawai'i 96720, USA}
\email{} 

\author[0000-0003-4793-7880]{Fabian Walter}
\affiliation{Max-Planck-Institut f\"ur Astronomie, K\"onigstuhl 17, D-69117, Heidelberg, Germany}
\email{} 

\author[0000-0002-7633-431X]{Feige Wang}
\affiliation{Department of Astronomy, University of Michigan, 1085 S. University Ave., Ann Arbor, MI 48109, USA}
\email{} 

\author[0000-0003-0643-7935]{Julien Wolf}
\affiliation{Max-Planck-Institut f\"ur Astronomie, K\"onigstuhl 17, D-69117, Heidelberg, Germany}  
\email{} 

\author[0000-0001-5287-4242]{Jinyi Yang}
\affiliation{Department of Astronomy, University of Michigan, 1085 S. University Ave., Ann Arbor, MI 48109, USA}
\email{} 

\begin{abstract}
We present three outliers on the quasar main sequence at $z>6$, as revealed by JWST/NIRSpec IFU observations from the \textsc{Aether} survey, a program targeting $\sim$200 quasars at cosmic dawn. The quasars, J0216$-$5226, J0320$-$3521, and J1429+5447, exhibit exceptionally strong \ion{Fe}{2} emission and weak H$\beta$ and [\ion{O}{3}] lines, indicating rapid chemical enrichment within the first Gyr of cosmic history. Their \ion{Fe}{2} strengths ($R_{\text{\ion{Fe}{2}},\lambda4570} \equiv \text{EW}_{\text{\ion{Fe}{2}}}/\text{EW}_{\text{H}\beta}$) range between $\sim2.5-3.6$, placing them among the top $\sim3$\% of SDSS quasars at ($0\lesssim z\lesssim 1$). The H$\alpha$ emission lines of J0216$-$5226 and J1429+5447 imply black hole masses of $\sim1-7\times10^9$ M$_{\odot}$ with $\lambda_{\text{Edd}}\sim0.1-0.6$, comparable to typical high-$z$ quasars, although virial black hole masses may be systematically overestimated in strong \ion{Fe}{2} emitters. In contrast, J0320$-$3521 hosts a lower mass black hole of $\sim0.2-2\times10^9$ M$_{\odot}$, suggesting possible super-Eddington rate accretion with $\lambda_{\text{Edd}}\sim0.5-6.0$.
The IFU data of J0320$-$3521 reveal a distinct emission-line region at a projected distance of $\sim1.5$ kpc, consistent with either an interacting companion or ionized gas associated to the quasar. We further identify a complex environment around J1429+5447, including two candidate companion galaxies or extended ionized gas structures.
These quasars with bright emission-line regions, J0320$-$3521 and J1429+5447, are radio-loud sources, while J0216$-$5226 remains undetected in radio ($R_{5\text{GHz,} \,3\sigma}\lesssim4$).
Together, these results confirm previous findings that highly enriched, rapidly growing black holes were already in place at cosmic dawn. Despite the small sample size, our results further highlight the diversity in environment and radio-jet properties among quasars that otherwise share the same extreme region of the quasar fundamental plane.
\end{abstract}


\section{Introduction} 
\label{sec:intro}

Quasars are supermassive black holes (SMBHs) undergoing rapid accretion at the centers of massive galaxies. Due to variations in their physical conditions, they exhibit a wide diversity of observational properties across the electromagnetic spectrum (see reviews by \citealt{Antonucci1993ARA&A..31..473A, Urry1995PASP..107..803U, Padovani2017A&ARv..25....2P, Inayoshi2020ARA&A..58...27I}). Previous studies have found well-defined trends that unify the diversity of quasars onto a quasar main sequence. One important trend, known as ``Eigenvector 1'' (or EV1), shows an anti-correlation between the strength of broad \ion{Fe}{2} emission and the average [\ion{O}{3}]$_{\lambda5007}$ strength (equivalent width, EW), and full-width-at-half-maximum (FHWM) of H$\beta$ \citep{Boroson1992ApJS...80..109B}. The primary driver of this relation is thought to be the level of accretion of matter onto the BH, i.e. the Eddington ratio, with higher Eddington ratios associated with weaker [\ion{O}{3}]$_{\lambda5007}$ emission and stronger \ion{Fe}{2} emission (e.g., \citealt{Boroson1992ApJS...80..109B, Sulentic2000ApJ...536L...5S, Marziani2001ApJ...558..553M, Dong2011ApJ...736...86D, Shen2014Natur.513..210S}). In addition, other factors, such as quasar orientation, have been proposed to influence the observed gas kinematics and emission line widths (e.g., \citealt{Marziani2001ApJ...558..553M, Risaliti2011MNRAS.411.2223R, Shen2014Natur.513..210S, Bisogni2017MNRAS.464..385B, Vietri2018A&A...617A..81V}). The quasar main sequence is therefore an empirical trend where quasars, when placed on a plot of H$\beta$ line width versus \ion{Fe}{2} strength ($R_{\text{\ion{Fe}{2}},\lambda4570} \equiv \text{EW}_{\text{\ion{Fe}{2}}}/\text{EW}_{\text{H}\beta}$), fall along a continuous, roughly triangular sequence rather than being randomly scattered. 

The relative \ion{Fe}{2} strength has been proposed as a proxy for broad-line region (BLR) metallicity \citep{Netzer2007ApJ...654..754N}, with photoionization models showing that $R_{\text{\ion{Fe}{2}}}$ increases with metallicity, although the response is sub-linear and degenerate with gas density and microturbulence (e.g., \citealt{Verner2003ApJ...592L..59V, Panda2018ApJ...866..115P}). In the rest-frame UV, the line ratio of \ion{Fe}{2}$_{\lambda2200-3090}$ and \ion{Mg}{2}$_{\lambda2800}$ can be used to study the chemical enrichment of the BLR, since there is an evolutionary time delay between the formation of \ion{Mg}{2} by core-collapse supernovae and \ion{Fe}{2} from type Ia supernovae within the galaxy. Even up to the highest redshift there has been no evidence for the evolution of \ion{Fe}{2}/\ion{Mg}{2} ratio (e.g., \citealt{DeRosa2011ApJ...739...56D, DeRosa2014ApJ...790..145D, Mazzucchelli2017ApJ...849...91M, Schindler2020ApJ...905...51S, Yang2021ApJ...923..262Y, Lai2022MNRAS.513.1801L}), which suggests that BLR are already chemically enriched within the first few hundred million years of star formation.

Within the quasar main sequence framework, quasars are traditionally divided into Population A (FWHM$_{\text{H}\beta} \leq 4000$ km s$^{-1}$) and B (FWHM$_{\text{H}\beta} > 4000$ km s$^{-1}$), roughly separating high- and low-accretors, respectively 
\citep{Sulentic2000ApJ...536L...5S, Sulentic2000ARA&A..38..521S}. The strongest $R_{\text{\ion{Fe}{2}}}$ emitters ($R_{\text{\ion{Fe}{2}}} > 1$) are classified as extreme Population A (called ``xA''), and are characterized by high metallicity and high accretion rates (e.g. \citealt{Negrete2012ApJ...757...62N, Marziani2014MNRAS.442.1211M, Panda2019ApJ...882...79P}).
The four dimensional (4D) Eigenvector 1 framework adds two more independent diagnostics including the centroid shift of the high ionization \ion{C}{4}$_{\lambda1549}$ line and the soft X-ray photon index ($\Gamma_{\text{soft}}$; e.g., \citealt{Sulentic2000ApJ...536L...5S, Sulentic2000ARA&A..38..521S, Marziani2001ApJ...558..553M}). 

Luminous quasars can have a dramatic impact on their surrounding gas through winds, outflows, heating, and jets. Their high-ionization emission lines often exhibit systematic blueshifts of up to several thousands of km s$^{-1}$, providing strong evidence for powerful ionized outflows (e.g., \citealt{Richards2002AJ....124....1R, Liu2026Natur.653..368L}). The powerful release of energy from the quasar can also induce outflows at kpc scales, which can influence the star formation activity in the host galaxy via active galactic nuclei (AGN) feedback, while also contributing to the self-regulation of SMBH accretion (e.g., \citealt{DiMatteo2005Natur.433..604D, Hopkins2016MNRAS.458..816H, Barai2018MNRAS.473.4003B}). About $\sim10-15$\% of the quasars furthermore launch powerful radio jets (e.g., \citealt{Blandford2019ARA&A..57..467B}), which does not seem to evolve significantly up to $z\sim6$ (e.g., \citealt{Banados2015ApJ...804..118B, Liu2021ApJ...908..124L, Gloudemans2021A&A...656A.137G, Diana2022MNRAS.511.5436D, Keller2024MNRAS.528.5692K}). Generally, quasars are classified as radio-loud if $R = F_{5\text{GHz}}/F_{4400\text{\AA}} >10$ in rest-frame \citep{Kellermann1989AJ.....98.1195K}. These quasar jets interact with the interstellar medium (ISM) and can both suppress and trigger star formation in their galaxies by heating the circumgalactic gas and compressing dense clouds, thereby directly influencing the growth rate of the galaxy (e.g., \citealt{Silk1998A&A...331L...1S, Bower2006MNRAS.370..645B, Cresci2015A&A...582A..63C, Bischetti2024ApJ...970....9B, Walter2025ApJ...983L...8W}). 
Within Population B, the fraction of radio-loud quasars has been found to be higher than in Population A at low-$z$ \citep{Sulentic2003ApJ...597L..17S, Zamfir2008MNRAS.387..856Z}, which has been attributed to orientation effects of the BLR and jet (influencing the observed FWHM$_{\text{H}\beta}$ and radio emission) and potentially the absence of highly-accreting, very massive black holes (see e.g., \citealt{Fraix-Burnet2017FrASS...4....1F, Ganci2019A&A...630A.110G, Marziani2018FrASS...5....6M}). 

In recent years, the James Webb Space Telescope (JWST) has opened up a new window to study the high-$z$ Universe in unprecedented detail, including the physical conditions and environments of the first quasars through programs such as ASPIRE (e.g., \citealt{Yang2023ApJ...951L...5Y, Wang2023ApJ...951L...4W}), EIGER (e.g., \citealt{Kashino2023ApJ...950...66K}), GA-NIFS (e.g., \citealt{Ubler2023A&A...677A.145U}) and SHELLQs-JWST (e.g., \citealt{Ding2025ApJ...993...91D}). These studies confirmed that their black holes masses are indeed $\sim10^{7-10}$ M$_{\odot}$ using the H$\beta$ tracer, making use of relations derived from low-$z$ reverberation mapping experiments. The initial samples of high-$z$ quasars also seem to follow the EV1 relations defined by low-$z$ quasars and show a chemically enriched BLR (e.g., \citealt{Yang2023ApJ...951L...5Y, Loiacono2024A&A...685A.121L, Liu2025arXiv251106085L, Dominika2025arXiv251009753D}). Furthermore, many quasars exhibit ionized [\ion{O}{3}] outflows (e.g., \citealt{Marshall2023A&A...678A.191M, Liu2024ApJ...976...33L, Phillips2025arXiv251022403P}) and several show evidence of companion galaxies undergoing mergers with the quasar host (e.g., \citealt{Marshall2023A&A...678A.191M, Decarli2024A&A...689A.219D, Liu2025arXiv251106085L}), as expected from cosmological simulations of $z\sim6$ quasars (e.g., \citealt{DiMascia2021MNRAS.503.2349D, Zana2022MNRAS.513.2118Z}). 
In addition, studies of the SMBH mass--stellar mass relation at high redshift have yielded mixed results: some studies suggest possible redshift evolution (e.g., \citealt{Pensabene2020A&A...637A..84P, Yue2024ApJ...966..176Y}), while other studies argue that the observed offset may instead be driven by larger intrinsic scatter and/or observational biases at high redshift (e.g., \citealt{Silverman2025ApJ...995L..67S, Li2025ApJ...981...19L, Ziparo2026arXiv260304358Z}).
In general, these findings are based on relatively small samples of $\lesssim 25$ quasars, most of which contain fewer than 10 objects. 

A recent JWST survey program called \textsc{Aether} obtained Near-InfraRed Spectrograph integral field unit (NIRSpec IFU) observations of $\sim$200 quasars at $z\geq5.7$ (see \citealt{Farina2024jwst.prop.5645F}, Farina et al. in prep.), providing the ideal statistical sample to study the most massive and highly accreting black holes in the early Universe. The sensitivity and wavelength range of the NIRSpec IFU instrument provides the ability to observe the rest-frame optical properties of quasars that were only accessible below $z\lesssim3$ pre-JWST. The newly obtained quasar properties can therefore be directly compared to the well-established low-$z$ quasar population from large-scale ground-based spectroscopic surveys such as Sloan Digital Sky Survey (SDSS; e.g., \citealt{Shen2011ApJS..194...45S}) and the Dark Energy Spectroscopic Instrument survey (DESI; e.g., \citealt{Chaussidon2023ApJ...944..107C}). 
The scale of the \textsc{Aether} survey not only enables a systematic characterisation of the average properties of the high-$z$ quasar population, but also the identification of the most extreme outliers, which can be used to test the limits of our current theoretical understanding. Inspection of the spectra revealed several such remarkable cases: quasars exhibiting exceptionally strong \ion{Fe}{2} emission alongside virtually absent [\ion{O}{3}] emission, indicative of very high metal enrichment and accretion rates in the early Universe. Objects exhibiting similar, albeit less extreme, \ion{Fe}{2} emission features have recently been reported by \citealt{Yang2023ApJ...951L...5Y} at $6.5<z<6.8$ and \citealt{Wolf2026A&A...707A.299W} at $z=7.64$.

This paper reports on three such quasars from the \textsc{Aether} sample, identified as extreme outliers on the quasar main sequence. Two of these quasars, J0320$-$3521 and J1429+5447, discovered by \cite{Ighina2023MNRAS.519.2060I} (and independently by \citealt{Yang2024MNRAS.528.2679Y}) and \cite{Willott2010AJ....139..906W} respectively, are radio-loud quasars at $z>6$ with high radio-loudness of $R\sim100-200$ (e.g., \citealt{Wang2011ApJ...739L..34W, Frey2011VLBI, Shao2020GMRT, Khusanova2022A&A...664A..39K, Ighina2023MNRAS.519.2060I}). In contrast, J0216$-$5226, discovered by \cite{Yang2019AJ....157..236Y}, remains undetected in the radio.

This paper is structured as follows. In Section~\ref{sec:obs}, we describe the JWST/NIRSpec, HST/WFC3, and rest-frame UV ground-based observations. The subsequent spectral analysis and derived properties of the three quasars are presented in Section~\ref{sec:spec_analysis}. In Section~\ref{sec:outflow_signatures}, we discuss the companion and outflow signatures seen in the NIRSpec IFU cubes within $\sim$10 kpc of the quasars. In Section~\ref{sec:radio}, we discuss the radio properties of our sources and repeat the modelling of the radio spectrum of J1429+5447 with new data at an ultra-low frequency of 54 MHz. Finally, our results are summarized in Section~\ref{sec:summary}. All magnitudes presented throughout the paper are in the AB magnitude system, and we assume a $\Lambda$ CDM cosmology with H$_{0}$= 70 km s$^{-1}$ Mpc$^{-1}$, $\Omega_{M}$ = 0.3, and $\Omega_{\Lambda}$ = 0.7. At $z=6$, assuming this cosmology, an arcsec corresponds to a proper distance of $\sim$5.7 kpc. 

\section{Observations \& Data reduction}
\label{sec:obs}

\subsection{JWST/NIRSpec}
\label{subsec:nirspec_obs}

The JWST/NIRSpec spectra of the three high-$z$ quasars discussed in this work are part of the cycle 3 JWST survey program (\#5645) called \textsc{Aether}, which is a snapshot NIRSpec/IFU survey of high-redshift quasars using the G395H high-resolution grating in combination with the F290LP filter. The \textsc{Aether} survey comprises 204 quasar fields spanning the redshift range $z=5.7$--$7.0$, selected such that both the H$\alpha$ and the H$\beta$+\ion{O}{3} emission-line complex fall within the wavelength coverage of the NIRSpec IFU G395H/F290LP disperser--filter configuration. The targets span nearly three orders of magnitude in luminosity, with monochromatic absolute magnitudes ranging from M$_\mathrm{1450}\sim-28$ mag to $\sim-22$ mag. The parent sample consists of all quasars known at $z>5.7$ at the time the proposal was prepared \citep[i.e., in 2024; see][]{Fan2022arXiv221206907F}. As \textsc{Aether} was designed as a JWST survey program, targets were selected from this parent sample primarily based on availability within gaps in the Long Range Plan scheduling. Consequently, the observed sample is not expected to be subject to additional selection biases beyond those inherent to the currently known high-redshift quasar population.

The NIRSpec IFU observations provide 3\arcsec$\times$3\arcsec\, spectral cubes (0.1\arcsec$\times$0.1\arcsec\, spatial elements) ranging from 2.87-5.27 $\mu$m with $R\sim2700$, covering the prominent H$\beta$ (4862.68 \AA), [\ion{O}{3}] doublet (4960.3 \& 5008.24 \AA), and H$\alpha$ (6564.61 \AA) emission lines. There is a gap in the spectra between $\sim4.0-4.2$ $\mu$m caused by the separation between the NRS1 and NRS2 detectors.
The quasars were observed with an on-source exposure time of 45 minutes, using the \textsc{NRSIRS2} readout pattern with 9 Groups, 1 Integration, and 4 \textsc{SMALL CYCLING} dither. The data were reduced using the standard STScI JWST pipeline\footnote{https://jwst-pipeline.readthedocs.io/en/stable/} with the addition of customized steps, following the procedure outlined in \cite{Loiacono2024A&A...685A.121L} and \cite{Decarli2024A&A...689A.219D}. More detailed data reduction steps and survey details are outlined in Farina et al. in prep and Decarli et al. in prep. 

From the cleaned data cubes, we extract the 1D quasar spectra using a circular aperture of 0.3\arcsec\, (3 pixels) centered on the peak quasar emission. This aperture radius ensures proper sampling of the NIRSpec PSF, while minimizing the contamination from the background. A local background subtraction is applied in each wavelength channel using the median signal measured within a 1--2.5\arcsec\ annulus. To correct for aperture flux losses, we apply the wavelength-dependent corrections as determined by \cite{Loiacono2024A&A...685A.121L} from a high-$z$ quasar with the same observational setup. The flux correction fractions applied range from 0.9085 at 2.90 $\mu$m to 0.874 at 5.20 $\mu$m (see Fig.~A.1 in \citealt{Loiacono2024A&A...685A.121L}). The resulting extracted spectra are shown in Fig.~\ref{fig:NIRSpec_spectra}. 
All three quasars discussed in this work show remarkably weak H$\beta$ and [\ion{O}{3}] emission lines, and strong iron continuum features. Initially, these have been identified in the survey data by visual inspection, and subsequently, confirmed as $>2\sigma$ outliers in rest-frame optical \ion{Fe}{2} strength compared to the full \textsc{Aether} population (see Sect.~\ref{subsec:FeII_strength_results} for more details).  

\subsection{Rest-frame UV spectroscopy}
\label{subsec:rest_UV_spec}

J0216$-$5226 was originally discovered by \cite{Yang2019AJ....157..236Y} at $z=6.41\pm0.05$ using Magellan/LDSS3, and has later been observed by \cite{Bigwood2024MNRAS.529.3511B} using a fixed slit on Magellan/FIRE covering $\sim8200-24000$ \AA. As noted in \cite{Bigwood2024MNRAS.529.3511B}, the Magellan/FIRE observation of J0216$-$5226 has a low signal-to-noise ratio (S/N) with a lack of strong emission lines. However, they do report a detection of the \ion{Mg}{2} line. J0320$-$3521 at $z=6.13\pm0.05$ was originally discovered by \cite{Ighina2023MNRAS.519.2060I} using Gemini-South/GMOS, and initially selected by combining the Dark Energy Survey and Rapid ASKAP Continuum Survey (RACS; \citealt{McConnell2020PASA...37...48M}). This quasar was independently selected as a quasar candidate by \cite{Yang2024MNRAS.528.2679Y}. J0320$-$3521 has not yet been followed up in the observed-frame NIR. Finally, J1429+5447 was first identified by \cite{Willott2010AJ....139..906W} in the Canada-France High-$z$ Quasar Survey (CFHQS) with an estimated redshift of $z=6.21$. Since then, it has been followed up by \cite{Shen2019ApJ...873...35S} with Gemini/GNIRS to obtain its near-infrared spectrum covering the C\textsc{iv} and \ion{Mg}{2} emission lines. 

In this work, we reanalyse the rest-frame UV spectra of J0216$-$5226 and J1429+5447 to investigate their chemical enrichment and BLR conditions via the \ion{Fe}{2}/\ion{Mg}{2} ratio. With their NIRSpec data available, we estimated the intrinsic dust reddening of each quasar in three steps. First, we corrected the ground-based spectra for Galactic foreground extinction using the \citet{Schlegel1998ApJ...500..525S} dust map value at each source's coordinates and the \citet{Cardelli1989ApJ...345..245C} extinction law ($R_V = 3.1$), and rebinned the Galactic-corrected spectra onto a common grid of constant velocity width (200 km s$^{-1}$). Second, we placed the ground-based spectra on the same absolute flux scale as the JWST/NIRSpec data by anchoring the \citet{Selsing2016A&A...585A..87S} quasar composite template to the JWST spectrum at rest-frame 7000 \AA\ and to the reddest ground-based spectral region ($3000$–$3250$ \AA). Finally, we fitted for the intrinsic $E(B-V)$ by reddening the (rescaled) composite template using the SMC bar extinction curve of \citet{Gordon2003ApJ...594..279G} and comparing it to the observed spectrum blueward of rest-frame 3000 \AA, minimizing $\chi^2$ while excluding windows contaminated by strong emission lines. This yielded a modest $E(B-V) = 0.005$ for J1429+5447 and $E(B-V) = 0.021$ for J0216$-$5226. The dereddened rest-frame UV spectra are shown in Fig.~\ref{fig:MgII_spec_fitting}.

\subsection{HST WFC3 Imaging}
\label{subsection:hst_im}

J1429+5447 has also been imaged using the Wide-Field Camera 3 (WFC3) on HST (program ID: 12184, PI: Xiaohui Fan), including the F105W and F140W filter. The quasar was observed on 2011-08-11, with an on-source exposure time of $\sim20$ minutes. In this work, we use the calibrated data products delivered through MAST and make use of the standard data reduction pipeline for WFC3\footnote{The data have been retrieved on 21 Nov 2025.}. The resulting astrometry can be accurate to a level of 0.1\arcsec. Since the JWST/NIRSpec IFU data of the \textsc{Aether} program have been obtained without target acquistion, its astrometry is not reliable. We therefore aligned the NIRSpec IFU cube of J1429+5447 with the HST images, using a translational shift, which led to a correction of $\sim0.25$\arcsec\, in both RA and Dec. 

\section{Spectral analysis}
\label{sec:spec_analysis}

\begin{figure*}
    \centering
    \includegraphics[width=1.0\linewidth]{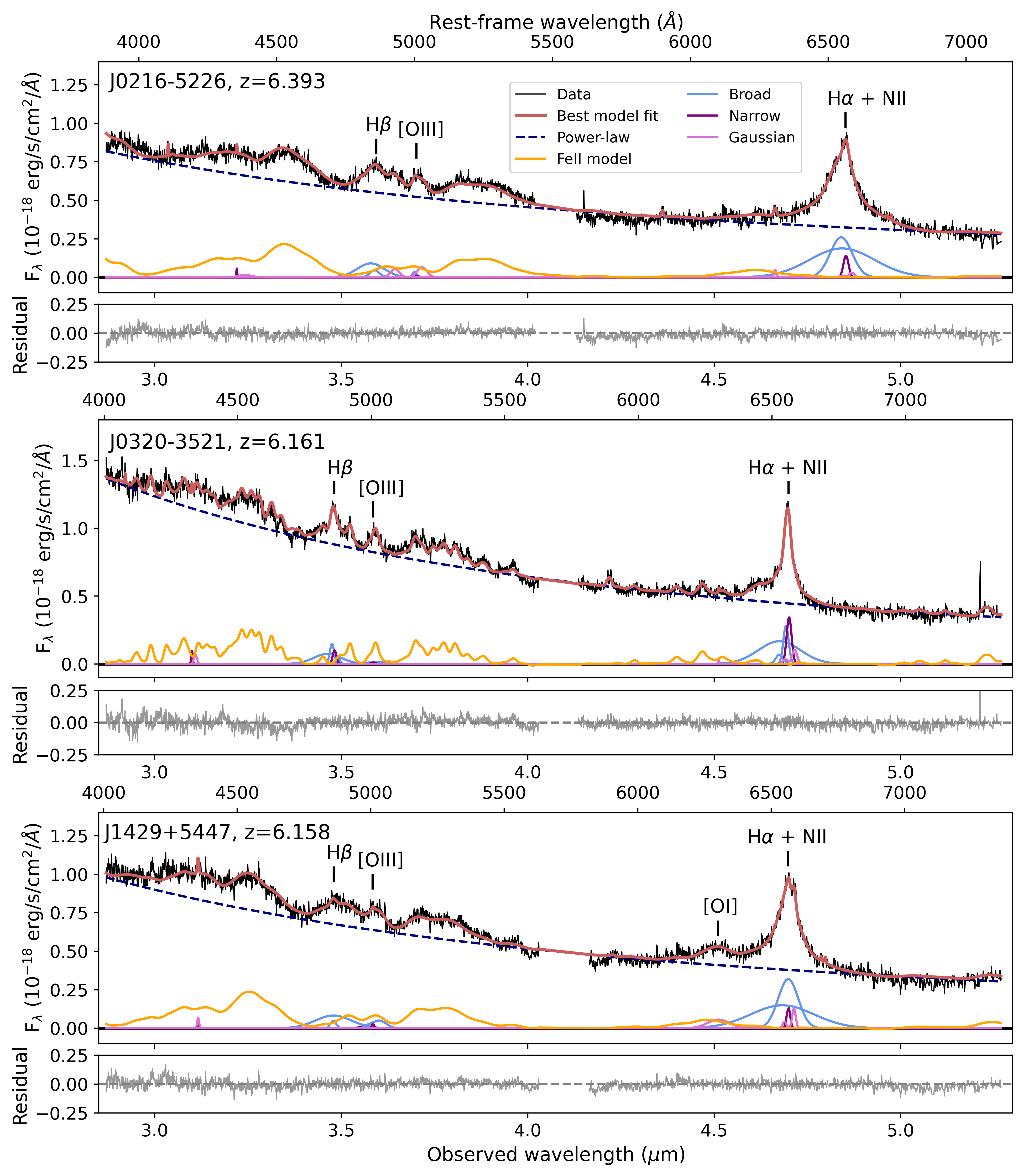}\vspace{-0.1cm}
    \caption{Extracted JWST/NIRSpec IFU spectra (F290LP filter/G395H grating) of J0216$-$5226 (top), J0320$-$3521 (middle), J1429+5447 (bottom). The best fit model is shown in red, consisting of a power law (dark blue), \ion{Fe}{2} continuum (yellow; combination of Mrk 493 and I Zw 1), and narrow (purple) and broad (light blue) emission lines. The residual spectrum after subtracting the model is shown in the panel below each spectrum. A zoom in of the H$\beta$+[\ion{O}{3}] complex using the two different \ion{Fe}{2} templates is shown in Appendix Fig.~\ref{fig:Hbeta_comparison_iron_templates}.}
    \label{fig:NIRSpec_spectra}
\end{figure*}

We model their spectra using the \textsc{python} package \textsc{Sculptor}\footnote{\url{https://sculptor.readthedocs.io/en/latest/index.html}} \citep{Schindler2022ascl.soft02018S}, which has been specifically developed for fitting high-$z$ quasar spectra. We fit both the quasar continuum and broad emission lines, and follow the general method used in numerous previous quasar studies (e.g., \citealt{DeRosa2014ApJ...790..145D, Shen2019ApJ...873...35S, Schindler2020ApJ...905...51S, Banados2021ApJ...909...80B, Farina2022ApJ...941..106F, Belladitta2025A&A...699A.335B}). The JWST/NIRSpec and Gemini/GNIRS and Magellan/FIRE spectra are fitted separately (while being similar in scale) to ensure an optimal continuum fit that is not affected by absolute flux calibration. 

The detailed fitting procedure of the JWST/NIRSpec \textsc{Aether} spectra is described in Protu\v{s}ov\'a et al.~in prep. In short, the quasar continuum is modeled by a power law component with slope $\alpha_{\lambda}$. This continuum model is fitted to the following rest-frame wavelength regions least contaminated by \ion{Fe}{2} and other emission lines, given by: 4140-4160~\AA, 4200-4230~\AA, 5450-5480~\AA, 5580-5630~\AA, 5730-5770~\AA, and 6900-7375~\AA. The iron pseudo-continuum is modeled using two different empirical \ion{Fe}{2} templates derived from narrow line Seyfert 1 galaxies: the I Zwicky-1 template from \cite{Boroson1992ApJS...80..109B}, which spans 3700-7480 \AA\,, and the Mrk 493 template from \cite{Park2022ApJS..258...38P}, which covers 4000-5600 \AA. The I Zw 1 template has been widely adopted in previous studies, enabling direct comparison with the literature, while the higher spectral precision of the Mrk 493 template generally provides improved fits to the data. Because the Mrk 493 template is limited to the rest-frame wavelength range 4000-5600 \AA, we use the I Zw 1 template for the remaining wavelength coverage. The \ion{Fe}{2} templates are fitted to rest-frame wavelength regions of 4435-4700~\AA\, and 5100-5550~\AA, and the broadening of the FWHM of the iron templates is allowed to vary up from 1000 to 5000 km s$^{-1}$. 
The continuum is modeled in two stages. In the first stage, the power-law continuum and the \ion{Fe}{2} template are each fit independently within the quoted spectral windows. The best-fit parameters from this stage serve as initial values for the second stage, in which the power-law continuum and \ion{Fe}{2} template are fit simultaneously. The resulting best-fit continuum model is then subtracted from the spectrum before fitting the emission lines.

The emission lines are fitted using multiple Gaussian components. In particular, the H$\beta_{\lambda4862.68}$ line is modeled using 2 or 3 broad components ($1000 < \text{FWHM} < 15000$ km s$^{-1}$) and 1 narrow component ($250 < \text{FWHM} < 1000$ km s$^{-1}$). The redshifts of the narrow components are tied together during the fit.
The [\ion{O}{3}] doublet is modeled with a single broad and narrow component, with the flux ratio of both narrow and broad line components of [\ion{O}{3}]$_{\lambda5008}$ and [\ion{O}{3}]$_{\lambda4960}$ fixed to 2.98. The FWHM of the [\ion{O}{3}] broad component is only allowed to vary between 1000-5000 km s$^{-1}$. 
Furthermore, H$\alpha_{\lambda6564}$ is fitted using 2 or 3 broad components and one narrow component as well, including the [\ion{N}{2}] doublet, which is modeled using 2 single Gaussians with a fixed ratio of [\ion{N}{2}]$_{\lambda6548}$ and [\ion{N}{2}]$_{\lambda6584}$ of 3. Additionally, the detected [\ion{O}{3}]$_{\lambda4363.21}$ and [\ion{O}{1}]$_{\lambda6302.046}$ emission lines are modeled with single Gaussians.
The number of broad components used for H$\alpha$ and H$\beta$ in the final fit is determined by selecting the model with the lowest Bayesian Information Criterion (BIC) value given by
\begin{align}
\label{eq:bic}
    \text{BIC} = \chi^2 - d_f\log(N)
\end{align}
with $\chi^2$ the chi-square statistic, $d_f$ the degrees of freedom, and N the sample size. For both \ion{Fe}{2} templates, only J0320$-$3521 requires 3 broad components, while 2 broad components suffice for J0216$-$5226 and J1429+5447.

In addition, the spectrum of J0216$-$5226 shows the presence of broad \ion{He}{1} emission lines at 4923\AA\, and 5017\AA, which can mimic an asymmetrical redshifted broad wing in H$\beta$ \citep{Veron2002A&A...384..826V}. This excess emission in the H$\beta$ broad wing has been named the ``red shelf'' or the ``shelf'' feature \citep{1985PASP...97..734M}. Therefore, to accurately model the H$\beta$ and [\ion{O}{3}] complex, we included a Gaussian model for both \ion{He}{1} lines. The inclusion of these \ion{He}{1} emission lines did not improve the fit of J0320$-$3521 and J1429+5447, and were therefore only included for J0216$-$5226. 

The best \textsc{Sculptor} fits are shown in Fig.~\ref{fig:NIRSpec_spectra}, and their derived physical parameters are summarized in Table~\ref{tab:sculptor_results}. From the H$\beta$, [\ion{O}{3}], and H$\alpha$ narrow line components, we derive redshifts of $z=6.393\pm0.001$ for J0216$-$5226, $z=6.161\pm0.002$ for J0320$-$3521, and $z=6.158\pm0.001$ for J1429+5447. The reported FWHM of H$\alpha$ and H$\beta$ are derived from the sum of their broad line components. 
The [\ion{O}{3}] and H$\beta$ emission lines of all three quasars are weak with rest-frame equivalent widths (EW$_{0}$) between 0.7-6.0~\AA\, in rest-frame for [\ion{O}{3}] and 15-23~\AA\, for H$\beta$. The Balmer decrement, defined as the flux ratio of H$\alpha$/H$\beta$, is 6.2$\pm$0.2 (8.4$\pm$0.3) for J0216$-$5226, 3.2$\pm$0.2 (3.6$\pm$0.1) for J0320$-$3521, and 4.8$\pm$0.2 (7.1$\pm$0.2) for J1429+5447, with the values obtained from using purely the \cite{Boroson1992ApJS...80..109B} iron template in brackets. These values are slightly higher than the typical ratio of $\sim$3 \citep{Dong2008MNRAS.383..581D} for 2 out of 3 quasars, suggesting possible dust attenuation.
However, the best fitting power law slopes of $-2.3 \lesssim \alpha_{\lambda} \lesssim -1.7$ are quite blue and similar to the unobscured type I quasar population of $-2 \lesssim \alpha_{\lambda} \lesssim -1$ \citep{VandenBerk2001AJ....122..549V}, which suggests these are not affected by dust extinction. The blue continuum slopes suggest that the weak H$\beta$ is unlikely to be caused solely from foreground dust extinction. This is not entirely unexpected, as the Balmer decrement does not necessarily provide a direct measure of the line-of-sight dust extinction, given that the observed Balmer emission originates in the BLR, where radiative transfer and optical depth effects can also influence the line ratios (e.g., \citealt{Dong2008MNRAS.383..581D, 2016MNRAS.461.4227H, Wu2023ApJ...950..106W}). The BLR may also be partially obscured by complex or patchy dust geometries (e.g., \citealt{Hodge2015ApJ...798L..18H, Gaskell2018MNRAS.478.1660G}).
Furthermore, we note that only J1429+5447 shows a prominent broad [\ion{O}{1}]$_{\lambda6302}$ line in its spectrum, which suggests the presence of an extended partially ionized gas region, and is often associated with radiative shocks (e.g., \citealt{Allen2008ApJS..178...20A}).  

\begin{deluxetable*}{lccccccc}
\tablewidth{0pt}
\tablecaption{Spectral fitting results and derived physical parameters \label{tab:sculptor_results}}
\tabletypesize{\footnotesize}
\tablehead{
\colhead{}
& \multicolumn{2}{c}{J0216$-$5226}
& \multicolumn{2}{c}{J0320$-$3521}
& \multicolumn{2}{c}{J1429+5447}
& \colhead{Unit} \\
\colhead{}
& \colhead{{\footnotesize P+22 \& BG92}} & \colhead{{\footnotesize BG92}}
& \colhead{{\footnotesize P+22 \& BG92}} & \colhead{{\footnotesize BG92}}
& \colhead{{\footnotesize P+22 \& BG92}} & \colhead{{\footnotesize BG92}}
& \colhead{}
}
\startdata
Redshift
& $6.393\pm0.001$ & $6.392\pm0.001$
& $6.161\pm0.002$ & $6.154\pm0.001$
& $6.158\pm0.001$ & $6.159\pm0.001$
& ---\\
FWHM$_{\text{H}\alpha}$ & $6074^{+103}_{-102}$ & $6274^{+108}_{-94}$ & $2219^{+324}_{-345}$ & $3597^{+162}_{-161}$ & $5207^{+148}_{-152}$ & $5303^{+130}_{-131}$ & km s$^{-1}$ \\
L$_{\text{H}\alpha}$ & $2.76\pm0.08$ & $2.96^{+0.06}_{-0.07}$ & $1.29^{+0.07}_{-0.06}$ & $1.54\pm0.02$ & $2.25^{+0.05}_{-0.06}$ & $2.48^{+0.04}_{-0.05}$ & $10^{44}$ erg s$^{-1}$ \\
FWHM$_{\text{H}\beta}$ & $5260^{+227}_{-189}$ & $4939^{+200}_{-172}$ & $1738^{+367}_{-208}$ & $2009^{+91}_{-81}$ & $5716^{+502}_{-585}$ & $4820^{+235}_{-255}$ & km s$^{-1}$ \\
EW$_{0,\text{H}\beta}^{(1)}$ & $23\pm1$ & $19\pm1$& $15\pm1$ & $17\pm1$ & $23\pm1$ & $17\pm1$ & \AA \\
EW$_{0,\text{\ion{Fe}{2}}}^{(2)}$ & $63\pm1$ & $69\pm1$ & $39\pm1$ & $47\pm1$ & $56\pm1$ & $59\pm1$ & \AA \\
$R_{\text{\ion{Fe}{2}}}$ & $2.72^{+0.08}_{-0.07}$ & $3.58^{+0.09}_{-0.08}$ & $2.56^{+0.15}_{-0.17}$ & $2.81^{+0.07}_{-0.04}$ & $2.50\pm0.10$ & $3.47^{+0.12}_{-0.07}$ & --- \\
EW$_{0,\text{[OIII]}}^{(1)}$ & $1.6\pm0.2$ & $1.1\pm0.1$ & $1.6\pm0.2$ & $0.7\pm0.1$ & $6.0\pm0.4$ & $1.9\pm0.2$ & \AA \\[1ex] \hline
L$_{\text{5100\AA}}$ & $1.73\pm0.01$ & $1.68\pm0.01$ & $2.40\pm0.01$ & $2.32\pm0.01$ & $1.87\pm0.01$ & $1.82\pm0.01$ & $10^{42}$ erg s$^{-1}$ \\
L$_{\text{bol}}^{(3)}$ & $8.15\pm0.01$ & $7.92\pm0.01$ & $11.34\pm0.01$ & $10.96\pm0.01$ & $8.82\pm0.02$ & $8.60\pm0.02$ & $10^{46}$ erg s$^{-1}$ \\
M$_{\text{BH,H}\alpha\text{,GH05}}$ & $1.81^{+0.61}_{-0.45}$ & $2.01^{+0.68}_{-0.49}$ & $0.15^{+0.09}_{-0.06}$ & $0.45^{+0.16}_{-0.11}$ & $1.18^{+0.41}_{-0.30}$ & $1.29^{+0.44}_{-0.32}$ & $10^9$ M$_{\odot}$ \\
M$_{\text{BH,H}\alpha\text{,DB+25}}$ & 6.6$\pm$0.2 & 7.2$\pm$0.2 & 0.68$\pm$0.17 & 1.7$\pm$0.1 & 4.3$\pm$0.2 & 4.8$\pm$0.2 & $10^9$ M$_{\odot}$ \\
M$_{\text{BH,H}\beta\text{,Shen+24}}$ &  $1.8^{+0.2}_{-0.1}$ & $1.6\pm0.1$ & $0.24\pm0.06$ & $0.31\pm0.02$ & $2.3\pm0.5$ & $1.6\pm0.2$ & $10^9$ M$_{\odot}$ \\
$\lambda_{\text{Edd,GH05}}$ & $0.36\pm0.01$ & $0.31\pm0.01$ & $6.0\pm1.9$ & $2.0\pm0.2$ & $0.60\pm0.04$ & $0.53\pm0.03$ & --- \\
$\lambda_{\text{Edd,DB+25}}$ & 0.10$\pm$0.01 & 0.09$\pm$0.01 & 1.3$\pm$0.3 & 0.50$\pm$0.04 & 0.16$\pm$0.01 & 0.14$\pm$0.01 & --- \\
$\lambda_{\text{Edd,Shen+24}}$ & $0.35\pm0.01$ & $0.39\pm0.01$ & $3.8\pm0.9$ & $2.80\pm0.2$ & $0.31\pm0.06$ & $0.43\pm0.04$ & --- \\[1ex] \hline
$z_{\text{\ion{Mg}{2}}}$ 
& \multicolumn{2}{c}{$6.307^{+0.005}_{-0.010}$}
& \multicolumn{2}{c}{---}
& \multicolumn{2}{c}{$6.110\pm0.005$}
& --- \\
FWHM$_{\text{\ion{Mg}{2}}}$
& \multicolumn{2}{c}{$4414^{+459}_{-460}$}
& \multicolumn{2}{c}{---}
& \multicolumn{2}{c}{$4125^{+1106}_{-594}$}
& km s$^{-1}$ \\
$F_{\text{\ion{Mg}{2}}}$
& \multicolumn{2}{c}{$20\pm2\times10^{-17}$} 
& \multicolumn{2}{c}{---}
& \multicolumn{2}{c}{$17\pm4\times10^{-17}$}
& erg s$^{-1}$ cm$^{-2}$ \\
$F_{\text{\ion{Fe}{2},2200-3090\AA}}$
& \multicolumn{2}{c}{$127\pm18\times10^{-17}$}
& \multicolumn{2}{c}{---}
& \multicolumn{2}{c}{$106\pm7\times10^{-17}$}
& erg s$^{-1}$ cm$^{-2}$ \\
$F_{\text{\ion{Fe}{2}}}/F_{\text{\ion{Mg}{2}}}$
& \multicolumn{2}{c}{$6.3^{+0.8}_{-0.8}$}
& \multicolumn{2}{c}{---}
& \multicolumn{2}{c}{$6.2^{+0.9}_{-1.2}$}
& --- \\
M$_{\text{BH,MgII}}$
& \multicolumn{2}{c}{$4.3^{+1.5}_{-1.2}\times10^9$}
& \multicolumn{2}{c}{---}
& \multicolumn{2}{c}{$3.2^{+3.3}_{-1.2}\times10^9$}
& M$_{\odot}$ \\
\enddata
\tablecomments{The quoted values in the left hand columns of each quasar have been derived using the combined \ion{Fe}{2} template from \cite{Park2022ApJS..258...38P} (at $<5600$~\AA) and \cite{Boroson1992ApJS...80..109B} (at $>5600$~\AA), and the values in the right hand columns using solely the \ion{Fe}{2} template from \cite{Boroson1992ApJS...80..109B}. The errors on the black hole masses and Eddington ratios only reflect the statistical errors, however, the systematic errors typically dominate with $\sim0.5$ dex. The rest-frame UV quantities (below the second horizontal line) do not depend on the choice of optical \ion{Fe}{2} template and are therefore listed spanning both columns for each source. (1) Rest-frame equivalent width (2) The \ion{Fe}{2} equivalent width obtained in wavelength range 4434--4684~\AA\, (3) Using the bolometric correction from \cite{Richards2006ApJS..166..470R}}
\vspace{-2em}
\end{deluxetable*}

\subsection{Rest-frame UV}

\begin{figure*}[ht]
    \centering
    \includegraphics[width=\textwidth]{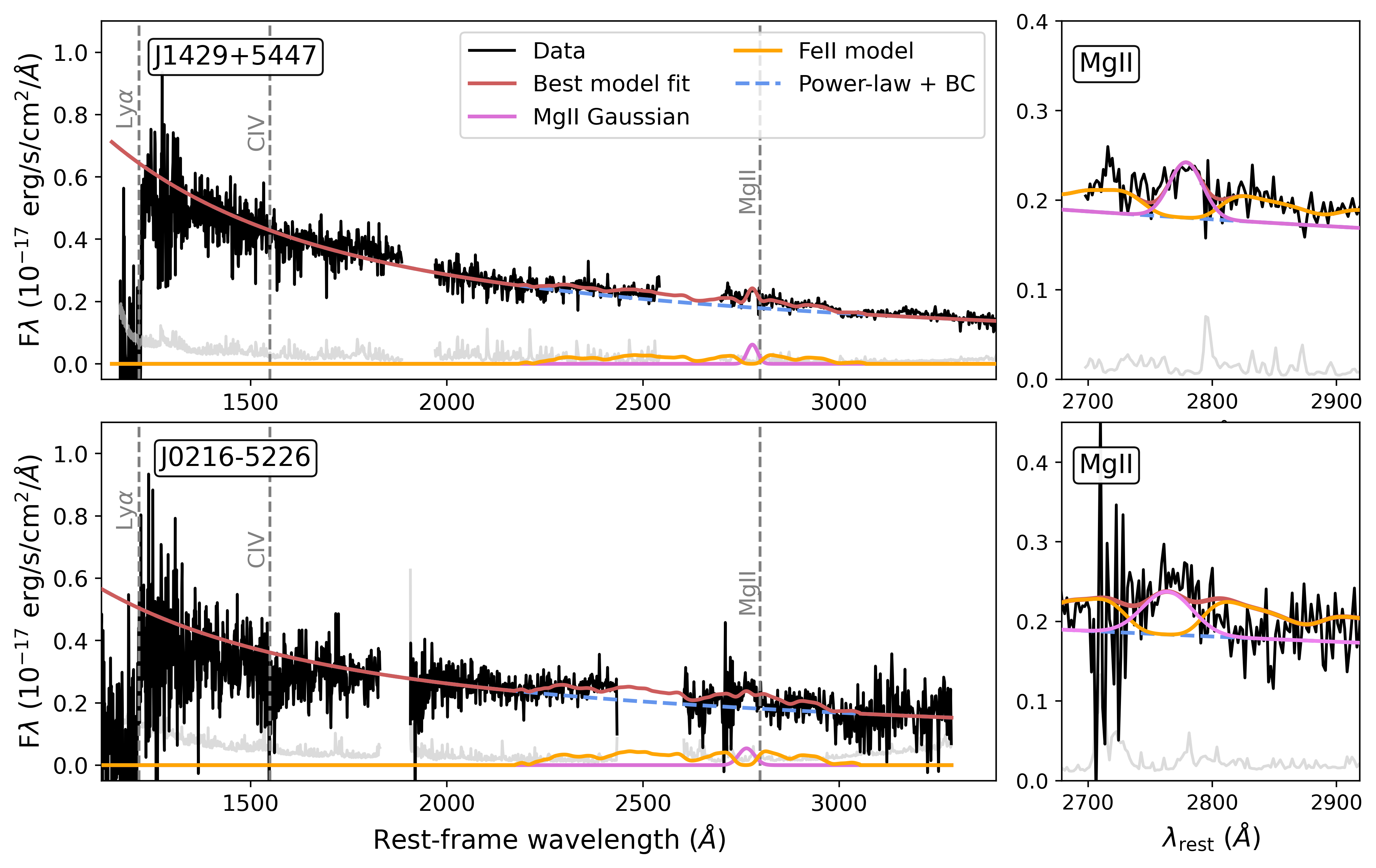}\vspace{-0.0cm}
    \caption{The dereddenend rest-frame UV spectra of J1429+5447 at $z=6.16$ (top panel) and J0216$-$5226 at $z=6.39$ (bottom panel) including best model fits. Top: The Gemini/GNIRS \citep{Shen2019ApJ...873...35S} of J1429+5447 showing the best model fit to the rest-frame UV and zoom-in on the \ion{Mg}{2} line (right panel). The error spectrum is given in grey. Bottom: The Magellan/FIRE spectra of J0216$-$5226 with best model fit. The errors on the UV spectrum (grey) are large, and therefore these fitting results are uncertain.}
    \label{fig:MgII_spec_fitting}
\end{figure*}

Furthermore, we repeat a similar fitting procedure for the rest-frame UV spectra of J0216$-$5226 and J1429+5447, using continuum regions adopted from \cite{Onoue2020ApJ...898..105O} ranging from 1275-1285~\AA, 1310-1325~\AA, 1425-1470~\AA, 1680-1710~\AA, 1975-2050~\AA\, in rest-frame. In addition, we add the wavelength region 3000-3100~\AA\, to better constrain the power-law slope redward of the \ion{Mg}{2} line. We utilize the empirical iron template from \cite{Vestergaard2001ApJS..134....1V} ranging from 2200-3500 \AA, and fix the FWHM value to 2500 km s$^{-1}$. The UV iron continuum is fitted and evaluated between 2200 and 3090~\AA\, rest-frame, and its redshift is fixed to the \ion{Mg}{2} line. The \ion{Mg}{2} emission line (at 2799.117~\AA) is modeled using a single Gaussian. The resulting best model fits are shown in Fig.~\ref{fig:MgII_spec_fitting} and their derived parameters again summarized in Table~\ref{tab:sculptor_results}. 
The rest-frame wavelength is calculated from the measured rest-frame optical redshifts, with the expected wavelengths of the Ly$\alpha$, \ion{C}{4}, and \ion{Mg}{2} lines indicated. The right panels of Fig.~\ref{fig:MgII_spec_fitting} show a zoom in on the \ion{Mg}{2} line and the best model fit. The rest-frame UV power-law slopes of $-1.70\pm0.02$ for J1429+5447 and $-1.41\pm0.04$ for J0216$-$5226 are similar to the rest-frame optical power-law slopes of $-1.93\pm0.02$ and $-1.77\pm0.02$, respectively. Any residual discrepancy is likely due to imperfect dust correction, compounded for J0216$-$5226 by the large flux errors in its Magellan/FIRE spectrum. This analysis could not be performed for J0320$-$3521, since there is no publicly available NIR data.

The \ion{Mg}{2} line models resulted in redshifts of $z_{\text{\ion{Mg}{2}}}=6.307\pm0.010$ for J0216$-$5226 and $z_{\text{\ion{Mg}{2}}}=6.110\pm0.005$ for J1429+5447, which indicates a blueshift (compared to the H$\alpha$ narrow components) of $-3480\pm400$ and $-2010\pm200$ km s$^{-1}$. These offsets are quite high compared to the average blueshift of $-480$ km s$^{-1}$ for $z\gtrsim6$ quasars found by \cite{Venemans2016ApJ...816...37V} and \cite{Decarli2018ApJ...854...97D} using [\ion{C}{2}]. However, the standard deviations of these distributions are large (e.g., 630 km s$^{-1}$ in \citealt{Venemans2016ApJ...816...37V}), and similar outliers are reported in those works. \\

\subsection{Measurement of the \ion{Fe}{2} strength and \ion{Fe}{2}/\ion{Mg}{2} ratio}
\label{subsec:FeII_strength_results}

The quasar fundamental plane relation EV1 established by \cite{Boroson1992ApJS...80..109B} demonstrates that a high \ion{Fe}{2} strength usually goes together with weak H$\beta$ and likely signifies a heavily accreting black hole. The rest-frame optical \ion{Fe}{2} strength is generally defined as $R_{\text{\ion{Fe}{2}},\lambda4570} \equiv \text{EW}_{\text{\ion{Fe}{2}}}/\text{EW}_{\text{H}\beta}$, with the \ion{Fe}{2} EW measured in the wavelength range of 4434-4684~\AA. We determine both of these equivalent widths from our \textsc{Sculptor} fit (see Fig.~\ref{fig:NIRSpec_spectra}), which results in \ion{Fe}{2} strengths of $3.58^{+0.09}_{-0.08}$ for J0216$-$5226, $2.81^{+0.07}_{-0.04}$ for J0320$-$3521, and $3.47^{+0.12}_{-0.07}$ for J1429+5447 in the case of using the I Zw 1 iron template from \cite{Boroson1992ApJS...80..109B}, which allows for direct comparison with literature values. These parameters are summarized in Table~\ref{tab:sculptor_results} for both the I Zw 1 and Mrk 493 templates, while Appendix Fig.~\ref{fig:Hbeta_comparison_iron_templates} compares the resulting fits and illustrates the differences between the template shapes.

\begin{figure}
    \centering
    \includegraphics[width=\columnwidth]{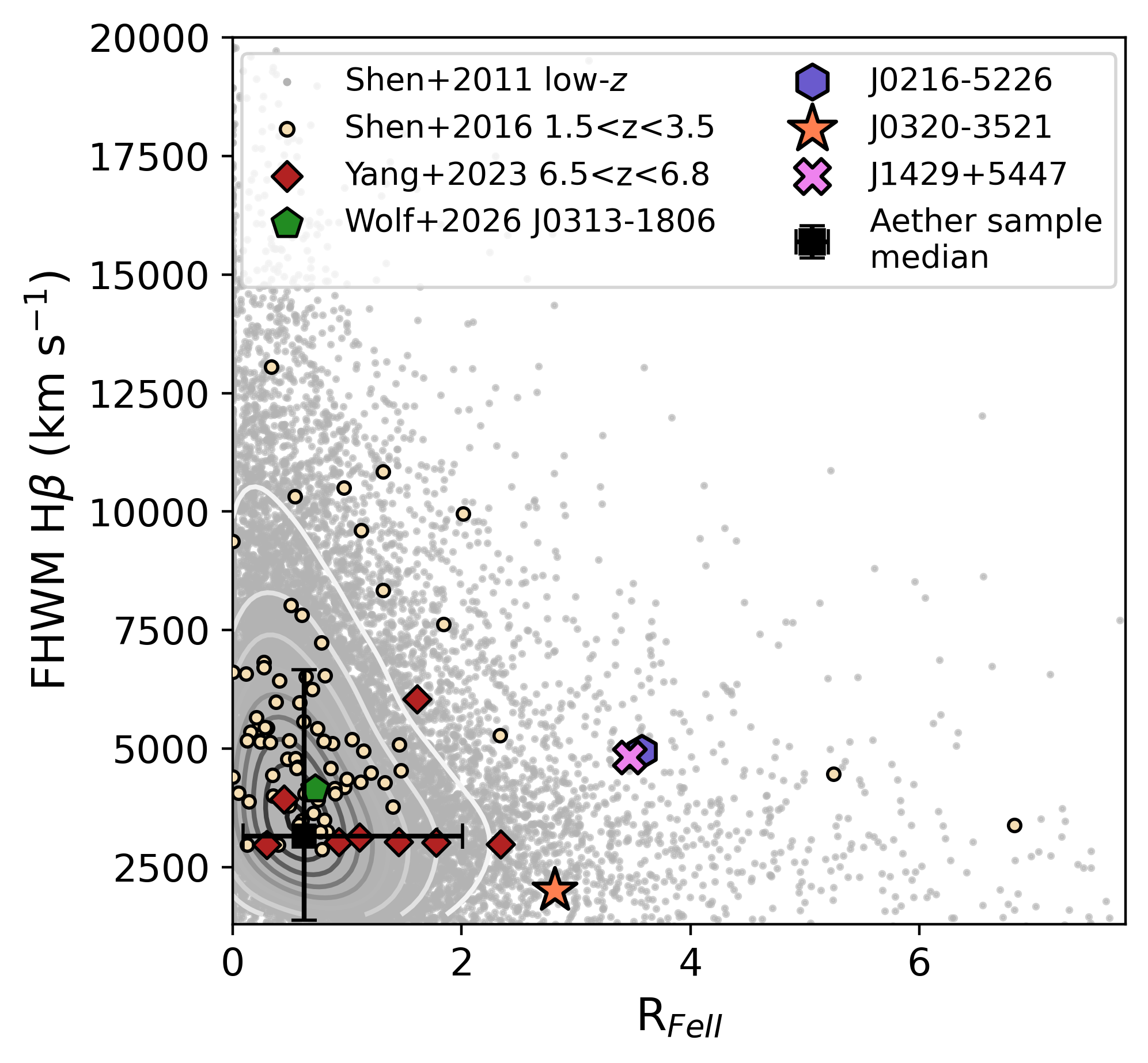}
    \caption{Eigenvector 1 plane of quasars with the \ion{Fe}{2} strength on the x-axis and FWHM of broad H$\beta$ line on the y-axis. J0216$-$5226, J0320$-$3521, and J1429+5447, are compared to lower redshift samples of \cite{Shen2011ApJS..194...45S} and \cite{Shen2016ApJ...817...55S}, and other high-$z$ quasars from \cite{Yang2023ApJ...951L...5Y} and \cite{Wolf2026A&A...707A.299W}. The median of the full \textsc{Aether} sample is shown in the black square with the error bars indicating the 2$\sigma$ sample distribution. The \ion{Fe}{2} strengths of our quasars are within the top 3\% of low-$z$ SDSS quasars and the highest amongst high-$z$ quasars observed so far.}
    \label{fig:FeII_FWHM_Hb}
\end{figure}

Fig.~\ref{fig:FeII_FWHM_Hb} compares the \ion{Fe}{2} strengths and FWHM$_{\text{H}\beta}$ of our quasars with the general quasar population from SDSS at $0 \lesssim z \lesssim 1$ by \cite{Shen2011ApJS..194...45S} and at $1.5<z<3.5$ by \cite{Shen2016ApJ...817...55S}. The grey contours indicate the density of low-$z$ quasars in this plane, with a median $R_{\text{\ion{Fe}{2}}}$ of 0.6. All three of our quasars have extremely high values of $R_{\text{\ion{Fe}{2}}}$ and are in the top $\sim3$\% of the low-$z$ quasar distribution, while being at $z\geq6$. The black square shows the median of the full \textsc{Aether} population (of $R_{\text{\ion{Fe}{2}}}\sim0.66$), with the error bars indicating the central 95\% ($\pm2\sigma$ equivalent) percentile of the distribution. The three quasars discussed in this work are $>2\sigma$ outliers of the \textsc{Aether} population and the only quasars with reliable measurements of $R_{\text{\ion{Fe}{2}}} \gtrsim3.0$.
Measurements from other recent high-$z$ quasar studies are also indicated in Fig.~\ref{fig:FeII_FWHM_Hb}, including the 8 quasars at $6.5<z<6.8$ from \cite{Yang2023ApJ...951L...5Y}, and J0313$-$1806 at $z=7.64$ from \cite{Wolf2026A&A...707A.299W}, again highlighting the exceptional $R_{\text{\ion{Fe}{2}}}$ measurements of our three quasars. We note that \cite{Wolf2026A&A...707A.299W} used the Mrk 493 iron template from \cite{Park2022ApJS..258...38P} to calculate $R_{\text{\ion{Fe}{2}}}$, which generally results in lower values ($\sim30\%$ in the most extreme cases) as also apparent in Table~\ref{tab:sculptor_results}. 

Quasar orientation and Eddington accretion ratio are believed to be the main physical drivers of the EV1 plane (e.g., \citealt{Boroson1992ApJS...80..109B, Sulentic2000ApJ...536L...5S, Marziani2001ApJ...558..553M, Shen2014Natur.513..210S, Marziani2018FrASS...5....6M}). For example, \cite{Shen2014Natur.513..210S} demonstrated that with increasing $R_{\text{\ion{Fe}{2}}}$, the average BH mass decreases and Eddington ratio increases, while the spread in FWHM$_{\text{H}\beta}$ at fixed $R_{\text{\ion{Fe}{2}}}$ is likely due to orientation (see also review by \citealt{Marziani2018FrASS...5....6M}).
The high $R_{\text{\ion{Fe}{2}}}$ values of our quasar imply these are in the extreme EV1 regime, which are generally associated with high Eddington accreters, making these excellent candidates for super-Eddington sources at $z>6$.
As expected from the trend observed at low-$z$, the [\ion{O}{3}] emission of our quasars is weak (see WLQ discussion in Sect.~\ref{subsec:WLQ}). 

Within the quasar main sequence framework, J0320$-$3521 is a typical extreme Population A source with $R_{\text{\ion{Fe}{2}}}> 1$ and FWHM$_{H\beta} < 4000$ km s$^{-1}$ (e.g. \citealt{Negrete2012ApJ...757...62N, Marziani2014MNRAS.442.1211M, Panda2019ApJ...882...79P}), however, J0216$-$5226 and J1429+5447 both exhibit a FWHM$_{H\beta} > 4000$ km s$^{-1}$. This region of the EV1 plane of high $R_{\text{\ion{Fe}{2}}}$ and FWHM$_{H\beta}$ is scarcely populated at low-$z$ as shown in Fig.~\ref{fig:FeII_FWHM_Hb}, and does not fall into a distinct sub-category. It is even unclear whether these $R_{\text{\ion{Fe}{2}}}$ values at low-$z$ are accurate: \citet{Sniegowska2018A&A...613A..38S} performed more detailed modeling and found that the quasars classified by \citet{Shen2011ApJS..194...45S} as having $R_{\text{\ion{Fe}{2}}} > 1.3$ generally dropped below this value once remeasured. These quasars are therefore extreme outliers in the EV1 plane. The \ion{Fe}{2} properties and fundamental plane relation of the full \textsc{Aether} sample will be published by Protu\v{s}ov\'a et al. (in prep.), which will provide a benchmark sample at high redshift.

Furthermore, with the rest-frame UV spectra of Gemini/GNIRS and Magellan/FIRE (see Fig.~\ref{fig:MgII_spec_fitting}), we can examine the \ion{Fe}{2}/\ion{Mg}{2} ratio for J0216$-$5226 and J1429+5447. We derive the \ion{Fe}{2} flux from 2200-3090~\AA\, in rest-frame using the \cite{Vestergaard2001ApJS..134....1V} iron template (see Sect.~\ref{sec:spec_analysis}) and the \ion{Mg}{2} flux from the emission line fit (see Fig. \ref{fig:MgII_spec_fitting}). This resulted in a \ion{Fe}{2}/\ion{Mg}{2} ratio of $6.3^{+0.8}_{-0.8}$ for J0216$-$5226 and $6.2^{+0.9}_{-1.2}$ for J1429+5447. These are within the same range (albeit on the high end) of the SDSS QSO sample from \cite{Calderone2017MNRAS.472.4051C} at $z\sim1-2$ and similar to higher-$z$ samples from \cite{DeRosa2011ApJ...739...56D}, \cite{Mazzucchelli2017ApJ...849...91M}, and \cite{Schindler2020ApJ...905...51S}, which all suggest a non-evolving \ion{Fe}{2}/\ion{Mg}{2} ratio in rapidly accreting quasars.

Finally, the velocity broadening and shift of \ion{Fe}{2} can be used to constrain the location of the \ion{Fe}{2} emitting region. Low-$z$ studies of SDSS quasars showed that the line width of \ion{Fe}{2} is significantly narrower than the broad component of H$\beta$, suggesting that \ion{Fe}{2} is produced further away from the central source than H$\beta$ (e.g., \citealt{Hu2008ApJ...687...78H, Hu2008ApJ...683L.115H}). This picture is further supported by direct reverberation mapping, which finds longer time lags for \ion{Fe}{2} than for H$\beta$ (e.g., \citealt{Barth2013ApJ...769..128B}), implying \ion{Fe}{2} originates from the outer BLR. Our fitting routine resulted in FWHM$_{\text{FeII}}$ values of 5000 km s$^{-1}$ for J0216$-$5226, 1421 km s$^{-1}$ for J0320$-$3521, and 4247 km s$^{-1}$ for J1429+5447 using the P22+BG92 \ion{Fe}{2} template. For J0320$-$3521 and J1429+5447, the FWHM$_{\text{FeII}}$/FWHM$_{\text{H}\beta}$ ratios of 0.82 and 0.74, respectively, are in agreement with the median value of $\sim$0.74 found by \cite{Hu2008ApJ...687...78H}.
For J0216$-$5226, we cannot constrain this ratio since the FWHM$_{\text{FeII}}$ maxes out at 5000 km s$^{-1}$. Further increasing the allowed FWHM only causes additional smoothing and blending of the \ion{Fe}{2} features, without improving the fit.

\subsection{Black hole mass and Eddington ratio}

We derive the black hole masses and Eddington ratios from both the H$\alpha$ and H$\beta$ emission lines using single epoch calibrations based on low-$z$ samples (e.g., \citealt{Greene2005ApJ...630..122G, Vestergaard2006ApJ...641..689V, Shen2011ApJS..194...45S, Shen2024ApJS..272...26S, DallaBonta2025A&A...696A..48D}). The H$\beta$ line typically provides the most reliable correlations \citep{Bentz2013ApJ...767..149B}, however, for the quasars in this work, the H$\beta$ emission line is weak, making the line fit less reliable than H$\alpha$. We therefore calculate both in this work for comparison. For the H$\alpha$-based BH masses, we adopt the scaling relations from both \cite{Greene2005ApJ...630..122G} and \cite{DallaBonta2025A&A...696A..48D} based on the H$\alpha$ line luminosity ($L_{\text{H}\alpha}$ in erg s$^{-1}$) and FWHM of the broad H$\alpha$ component given by
\begin{align}
\begin{split} 
\log_{10}\Big(\frac{M_{\mathrm{BH,H}\alpha}}{M_{\odot}}\Big)
    &= a + b(\log_{10}L_{\mathrm{H}\alpha} - 42) \\
    &\quad + c(\log_{10}\mathrm{FWHM}_{\mathrm{H}\alpha} - 3.5)
\end{split}
\end{align}
with $a=7.371$, $b=0.812$, and $c=1.634$ for \cite{DallaBonta2025A&A...696A..48D} and $a=7.331$, $b=0.55$, and $c=2.06$ for \cite{Greene2005ApJ...630..122G}. While the latter has been rederived from the relationship between H$\beta$ lag and 5100\AA\, continuum, it is often used in literature and therefore useful for comparison to previous work. The typical systematic uncertainty for these relations is $\sim0.4-0.5$ dex. For the H$\beta$ derived BH masses we use the relation from \cite{Shen2024ApJS..272...26S} for H$\beta$, given by
\begin{align} 
\begin{split}
\log_{10}\Big(\frac{M_{\mathrm{BH,H}\beta}}{M_{\odot}}\Big)
    &= 0.85 + 0.5\log_{10}\Big(\frac{L_{\lambda,5100}}{10^{44} \, \text{erg} \, \text{s}^{-1}} \Big) \\
    &\quad + 2.0\log_{10}\Big(\frac{\text{FWHM}_{\text{H}\beta}}{\text{km} \,\text{s}^{-1}}\Big)
\end{split}
\end{align}
with $L_{\lambda,5100}$, the monochromatic luminosity at 5100~\AA\, in rest-frame. These relations result in a wide range of BH mass estimates between $\sim2\times10^8$ M$_{\odot}$ and $\sim7\times10^9$ M$_{\odot}$ as shown in Table~\ref{tab:sculptor_results}. Specifically, the H$\alpha$-based BH masses derived from \cite{DallaBonta2025A&A...696A..48D} are $\sim1-7\times10^9$ M$_{\odot}$, similar to the general high-$z$ quasar population.
It is important to note that the reported errors are calculated by propagating uncertainties, however, in reality these are dominated by systematic uncertainties from the derived empirical relations, which are typically $\sim0.5$ dex. The derived BH mass values depend on the scaling relations and the \ion{Fe}{2} templates used, and therefore all values are quoted in Table~\ref{tab:sculptor_results}.

For comparison, we also derive BH mass estimates from the \ion{Mg}{2} lines, using the empirical relation from \cite{Shen2024ApJS..272...26S} given by
\begin{align}
\begin{split}
    \log_{10}\Big(\frac{M_{\mathrm{BH,MgII}}}{M_{\odot}}\Big) 
    &= -2.05 + 0.6\log_{10}\Big(\frac{L_{3000}}{10^{45} \text{erg}\,\text{s}^{-1}}\Big) \\
    &+ 3.0\log_{10}\Big(\frac{\text{FWHM}_{\text{MgII}}}{\text{km} \,\text{s}^{-1}}\Big)
\end{split}
\end{align}
with $L_{3000}$ the local continuum luminosity at 3000\,\AA\, in rest-frame. This yields BH masses of $4.3^{+1.5} _{-1.2}\times10^9$ $M_{\odot}$ for J0216$-$5226, and $3.2^{+3.3} _{-1.2}\times10^9$ $M_{\odot}$ for J1429+5447. These values are similar to the BH mass estimates derived from H$\alpha$ using the relation from \cite{DallaBonta2025A&A...696A..48D} (see Table~\ref{tab:sculptor_results}), and to the BH masses derived from \ion{Mg}{2} in previous work, such as $2.08\pm0.92\times10^9\, M_{\odot}$ for J0216$-$5226 by \cite{Bigwood2024MNRAS.529.3511B} and $2.35\pm0.36\times10^9\, M_{\odot}$ by \cite{Shen2019ApJ...873...35S}. However, the high blueshift of \ion{Mg}{2} with respect to H$\alpha$ found of $\sim-3000$ and $-2000$ km s$^{-1}$ for J0216$-$5226 and J1429+5447, respectively, suggests that non-virial motions, such as radiation-driven winds or outflows, might have impacted the measured line width, making the \ion{Mg}{2} based BH mass estimations less reliable \citep{Marziani2013ApJ...764..150M}. 

The bolometric luminosity $L_{\text{bol}}$ is derived from $L_{\lambda,5100}$ using the bolometric correction factor from \cite{Richards2006ApJS..166..470R}, resulting in $L_{\text{bol}} = 9.26\times\lambda L_{\lambda,5100}$. Finally, the Eddington luminosity $L_{\text{Edd}}$ is derived using
\begin{align}
\label{eq:ledd}
    L_{\text{Edd}} &= \frac{4\pi G m_p c M_{\text{BH}}}{\sigma_T} = 1.257\times10^{38}\times \Big(\frac{M_{\text{BH}}}{M_{\odot}}\Big)\, \text{erg}\, \text{s}^{-1}
\end{align}
with $G$ the gravitational constant, $m_p$ the proton mass, and $\sigma_T$ the Thomson cross-section of an electron. The resulting Eddington ratios ($\lambda_{\text{Edd}} = L_{\text{bol}}/L_{\text{Edd}}$) are highly dependent on the BH mass estimate used, suggesting values between $\sim0.09-0.39$ for J0216$-$5226, $\sim0.5-6.0$ for J0320$-$3521, and $\sim0.14-0.60$ for J1429+5447. These inferred Eddington ratios indicate that the quasars are efficiently accreting, with potentially super-Eddington accretion in the case of J0320$-$3521.
However, at face value, these Eddington ratios are not exceptional in the context of $z\gtrsim6$ quasars, which often have $\lambda_{\text{Edd}}\sim1$ (e.g., \citealt{Farina2022ApJ...941..106F}). Given their extreme Fe\textsc{ii} strength on the EV1 plane, these are expected to be among the highest accreting quasars. However, $R_{\text{\ion{Fe}{2}}}$ does not directly translate to accretion rate, as additional parameters such as BLR structure, orientation, and jet-related physics may also influence the observed spectral properties (see e.g., \citealt{Zamfir2008MNRAS.387..856Z, Marziani2009A&A...495...83M, Kollatschny2011Natur.470..366K, Shen2014Natur.513..210S}). On the other hand, applying standard single-epoch virial calibrations to extreme Fe\textsc{ii}-strong quasars remains uncertain and introduces systematic biases in the inferred black hole masses and Eddington ratios. Reverberation mapping observations show that highly accreting systems exhibit a relatively shorter time lag relative to the canonical $R-L$ relation, resulting in overestimated BH masses and underestimated accretion rates (e.g., \citealt{Du2014ApJ...782...45D, Du2019ApJ...886...42D, Maithil2022MNRAS.515..491M}). Using a high-$z$ quasar sample, \cite{Maithil2022MNRAS.515..491M} found this difference to be a factor $\sim2$ on average. The derived BH masses and accretion rates should therefore be interpreted with caution.

The possible super-Eddington nature of J0320$-$3521 has been previously presented in \cite{Ighina2025ApJ...990L..56I}, who observed remarkably high soft X-ray emission that is consistent with super-Eddington accretion and cannot be produced by relativistic jets. The JWST/NIRSpec observations presented here provide additional support for this interpretation, consistent with the efficient accretion expected for extreme sources on the fundamental plane.

One of the open questions in the field is how these massive BHs form and evolve into present-day systems. The extremely high BH masses of high-$z$ quasars provide strong constraints on BH formation models, since they must be assembled within the $\leq1$ Gyr available at $z\gtrsim6$ (see reviews by e.g., \citealt{Volonteri2012Sci...337..544V, Inayoshi2020ARA&A..58...27I, Trakhtenbrot021IAUS..356..261T, Fan2023ARA&A..61..373F}). The BH masses and Eddington ratios derived for J0216$-$5226, J0320 $-$3521, and J1429+5447 span a wide range, from $\sim10^{8-10}$ M$_{\odot}$ and $0.1-6.0$, reflecting both the intrinsic diversity of the sample and the systematic uncertainties inherent to single-epoch virial mass estimates. The subsequent growth of these BHs depends on the accretion rate, the radiative efficiency, and the fraction of cosmic time spent actively accreting, i.e. the quasar duty cycle. The latter is poorly constrained but likely small: proximity-zone and quasar-lifetime measurements at $z\sim6$ favor short episodes of luminous, unobscured accretion (e.g., \citealt{Eilers2020ApJ...900...37E, Morey2021ApJ...921...88M, Eilers2024ApJ...974..275E}), implying that the near-Eddington rates observed here cannot be sustained over the age of the Universe. Nonetheless, given plausible duty cycles and episodic accretion histories, the BHs studied in this work could feasibly grow into the most massive $\sim10^{10-11}$ M$_{\odot}$ BHs known locally (e.g., \citealt{McConnell2011Natur.480..215M, Thomas2016Natur.532..340T, Mehrgan2019ApJ...887..195M}).

\subsection{Weak-line quasar phenomenon}
\label{subsec:WLQ}

All three of the quasars discussed in this work have been previously identified as weak-line quasars (WLQ; \citealt{Diamond-Stanic2009ApJ...699..782D}) in their respective discovery papers and follow-up work (J0216$-$5226; \citealt{Bigwood2024MNRAS.529.3511B}, J0320$-$3521; \citealt{Ighina2023MNRAS.519.2060I}, J1429+5447; \citealt{Shen2019ApJ...873...35S}), based on their weak Ly$\alpha$ and \ion{C}{4} emission lines in the rest-frame UV. The physical origin of WLQs remains debated, with possible explanations including a gas-deficient or underdeveloped BLR (e.g., \citealt{Hryniewicz2010MNRAS.404.2028H, Plotkin2015ApJ...805..123P}) and a soft ionizing continuum possibly driven by super-Eddington accretion, which suppresses the strength of high-ionization emission lines (e.g., \citealt{Shemmer2008ApJ...682...81S, Leighly2007ApJS..173....1L, Meusinger2014A&A...568A.114M, Laor2011MNRAS.417..681L}).

The JWST/NIRSpec data presented in this work enable a comparison with their rest-frame optical emission-line properties. Previous work suggests that generally the low-ionization lines of WLQs, such as H$\beta$, H$\alpha$, and \ion{Mg}{2}, are not exceptionally weak compared to typical quasars (e.g., \citealt{Plotkin2015ApJ...805..123P, Chen2024ApJ...972..191C}), while their high-ionization lines, such as [\ion{O}{3}]$_{\lambda5007}$, are universally weak or absent. Also, stronger optical \ion{Fe}{2} emission appears to be a distinct feature of WLQs with $R_{\text{FeII}} \gtrsim1$ (e.g., \citealt{Shemmer2010ApJ...722L.152S, Plotkin2015ApJ...805..123P, Marziani2016Ap&SS.361...29M, Chen2024ApJ...972..191C}). 
Our quasars exhibit both strong \ion{Fe}{2} emission as well as extremely weak [\ion{O}{3}] emission lines, with EW$_{0,[\text{OIII}]}$ values of $1.6\pm0.2$ \AA\, for J0216$-$5226, $1.6\pm0.2$ \AA\, for J0320$-$3521, and $6.0\pm0.4$ \AA\, for J1429+5447. 
These findings naturally fit within the scenario of WLQs being extreme population A sources with high accretion rates, as suggested by \cite{Marziani2016Ap&SS.361...29M}. This view is supported by observed X-ray weakness consistent with potential absorption due to a puffed-up inner disk caused by super-Eddington accretion (e.g., \citealt{Luo2015ApJ...805..122L}). The X-ray properties of J0320$-$3521 and J1429+5447 are discussed in Sect.~\ref{sec:radio}.

\section{Quasar companion \& outflow signatures}
\label{sec:outflow_signatures}

\begin{figure*}
\centering 
{\includegraphics[width=1.0\textwidth]{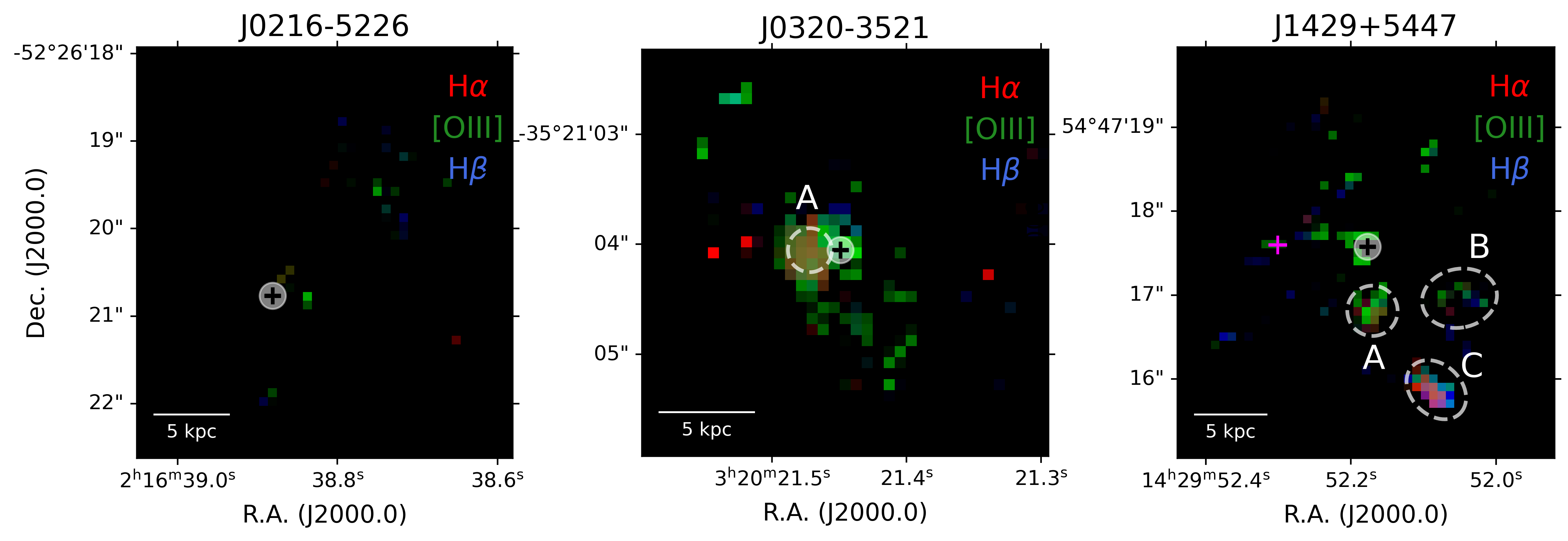}
}
\caption{\label{fig:rgb_maps} Composite H$\alpha$ (red), [\ion{O}{3}] (green), and H$\beta$ (blue) integrated intensity maps of J0216$-$5226 (left), J0320$-$3531 (middle), and J1429+5447 (right) after modeling and subtracting the quasar emission (showing only S/N$>1$). The quasar location is indicated with a black cross. The map of J0320$-$3531 (middle) shows a single clear emission line region to the East of the quasar, denoted A. The map of J1429+5447 shows three distinct emission line regions, which are highlighted by the circles and denoted A, B, and C. The pink cross on the J1429+5447 map shows the sky location of a previously observed CO(2-1) emission by \cite{Wang2011ApJ...739L..34W}. We interpret region C as a foreground galaxy, therefore its apparent emission originates from the galaxy continuum instead of the specific emission lines.}
\end{figure*}

The NIRSpec IFU cubes enable the detection of outflows and close companions within $\sim10$ kpc of the quasar (3\arcsec$\times$3\arcsec\, field of view). We therefore investigate integrated intensity maps centered on key rest-frame emission lines to search for such signatures. In particular, we construct maps around key rest-frame optical emission lines: H$\alpha$, H$\beta$, \ion{He}{2}, [\ion{N}{2}], [\ion{O}{1}], [\ion{O}{2}]$_{\lambda\lambda3726,3729}$, [\ion{O}{3}], [\ion{S}{2}]$_{\lambda\lambda6716,6731}$. 
These moment maps have been created for the full \textsc{Aether} sample through careful quasar subtraction and are presented by Decarli et al. (in prep). In short, we model the quasar point spread function based on the wings of the broad H$\alpha$ and H$\beta$ emission, which are dominated by the nuclear emission. The PSF model is scaled to match the quasar spectrum in the central pixels, and subtracted off the cubes. We create masks for the various emission lines based on the 3D structure of the [\ion{O}{3}], H$\alpha$, and H$\beta$ emission, and inferred moment-0 (integrated intensity), moment-1 (velocity field), and moment-2 (velocity dispersion) maps by collapsing the masked cubes after shifting the masks to the observed wavelength of each transition at the redshift of the quasar. Further details of the methodology will be presented in Decarli et al. (in prep). The composite H$\alpha$ (red), [\ion{O}{3}] (green), and H$\beta$ (blue) integrated intensity maps of the three quasars in this work are shown in Fig.~\ref{fig:rgb_maps}. These three emission lines are the strongest tracers of these high-$z$ quasar environments, with [\ion{O}{3}] kinematics revealing ionized outflows and both [\ion{O}{3}] and the Balmer lines sensitive to the presence of companion galaxies (see also e.g., \citealt{Wang2023ApJ...951L...4W, Kashino2023ApJ...950...66K, Decarli2024A&A...689A.219D}). In these maps, all pixels with S/N$\leq1$ are masked for presentation purposes to highlight the detected emission line regions and reduce random noise. We note that these emission line maps are constructed from narrow spectral windows centered on the wavelengths expected for each line at the quasar's redshift, and can therefore include both genuine line emission from sources at (or near) the quasar's redshift and continuum flux from unrelated sources at any redshift that happen to fall within the same wavelength range.

\subsection{Companion galaxy or ionized gas region of J0320$-$3521}

\begin{figure}
\centering 
{\includegraphics[width=\columnwidth]{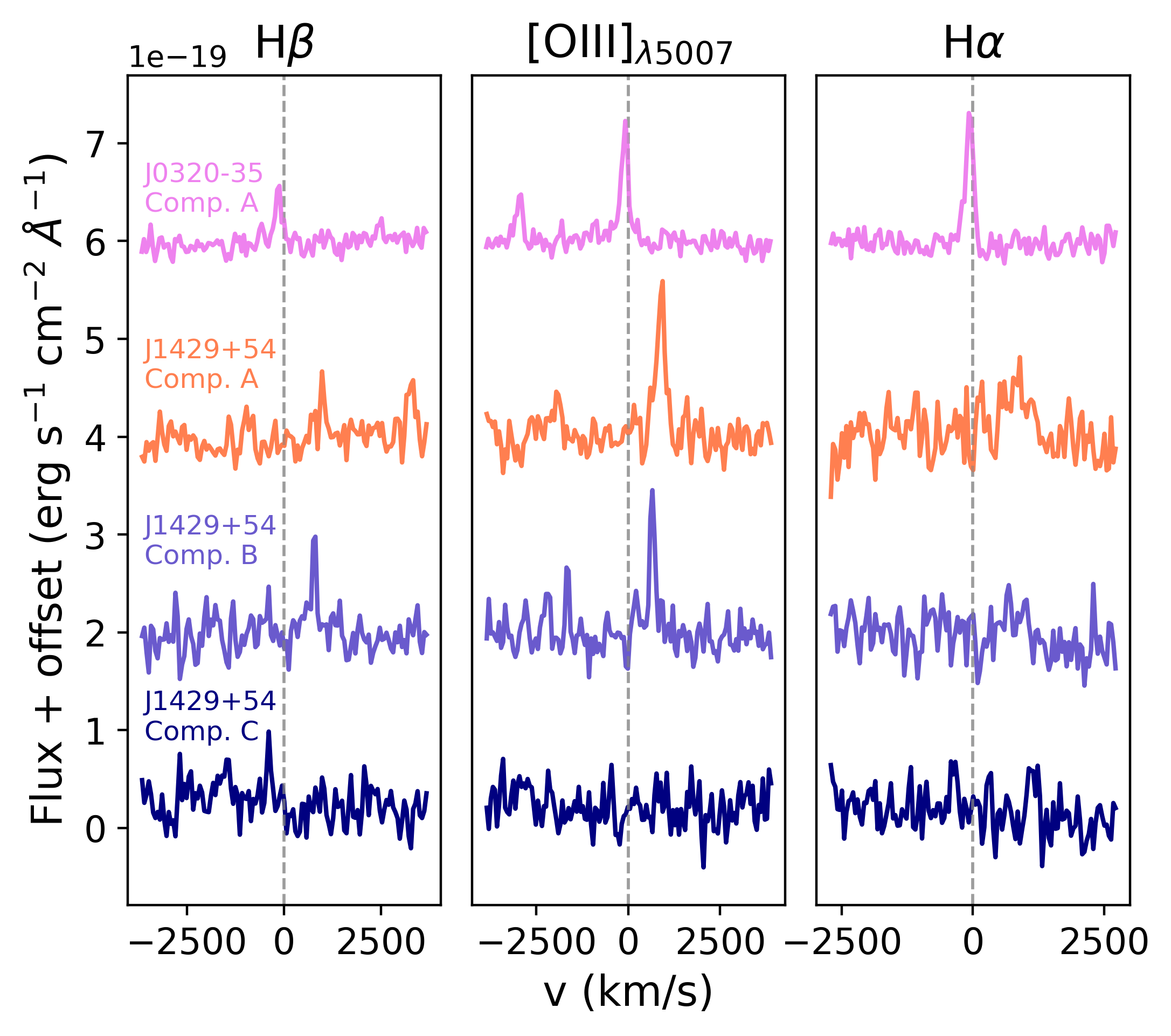}
}
\caption{\label{fig:IFU_companions_spectra} Optimally extracted emission-line spectra from region A (pink) in the IFU cube of J0320$-$3521 and region A (orange), B (purple), and C (blue) in J1429+5447. Spectra are vertically offset for visualization purposes. The panels show zoom-ins on H$\beta$, [\ion{O}{3}]$_{\lambda5007}$, and H$\alpha$. Grey dashed lines (at $\text{v}= 0$) mark the expected wavelength based on the quasar redshifts.}
\end{figure}

There is a clear detection of a single emission line region $\sim1.5$\,kpc away from J0320$-$3521 ($\sim$0.269\arcsec\, East and $\sim$0.0087\arcsec\, North, see middle panel of Fig.~\ref{fig:rgb_maps}). 
To investigate its nature, we extract its spectrum from the PSF-subtracted IFU cube by placing an aperture of 0.2\arcsec\, on the emission line region (at RA 03:20:21.47, Dec -35:21:04.06), chosen to optimize the S/N. The detected emission lines are shown in Fig.~\ref{fig:IFU_companions_spectra} (pink line) with clear detections of H$\beta$, [\ion{O}{3}]$_{\lambda4960}$, [\ion{O}{3}]$_{\lambda5007}$, and H$\alpha$. All of these emission lines have narrow widths of $\sim$200 km s$^{-1}$ and are slightly blueshifted by $\sim60-150$ km s$^{-1}$ relative to the H$\alpha$ emission line of J0320$-$3521. The emission lines \ion{He}{2}, \ion{O}{3}$_{\lambda4363}$, [\ion{N}{2}]$_{\lambda6583}$, [\ion{O}{1}]$_{\lambda6300}$, and [\ion{S}{2}$_{\lambda\lambda6716,6731}$] are not detected in this region. Due to the non-detection of [\ion{N}{2}]$_{\lambda6583}$, we place an upper limit on the [\ion{N}{2}]/H$\alpha$ ratio of 0.27. Combined with an [\ion{O}{3}]/H$\beta$ ratio of 2.5, the location on the BPT diagram of \cite{Kewley2006MNRAS.372..961K} does not uniquely distinguish between excitation from star formation in a companion galaxy, and ionization from AGN-associated outflowing or inflowing gas. The small velocity offset and the narrow line widths are consistent with both a mild kinematic offset from a companion galaxy and low-velocity ionized gas. This is also apparent from the [\ion{O}{3}]$_{\lambda5007}$ moment maps shown in Fig.~\ref{fig:OIII_maps}, with similar kinematics seen in H$\beta$ and H$\alpha$ (see Appendix Fig.~\ref{fig:Hb_maps} and \ref{fig:Ha_maps}).

\begin{figure*}
\centering 
{\includegraphics[width=1.0\textwidth]{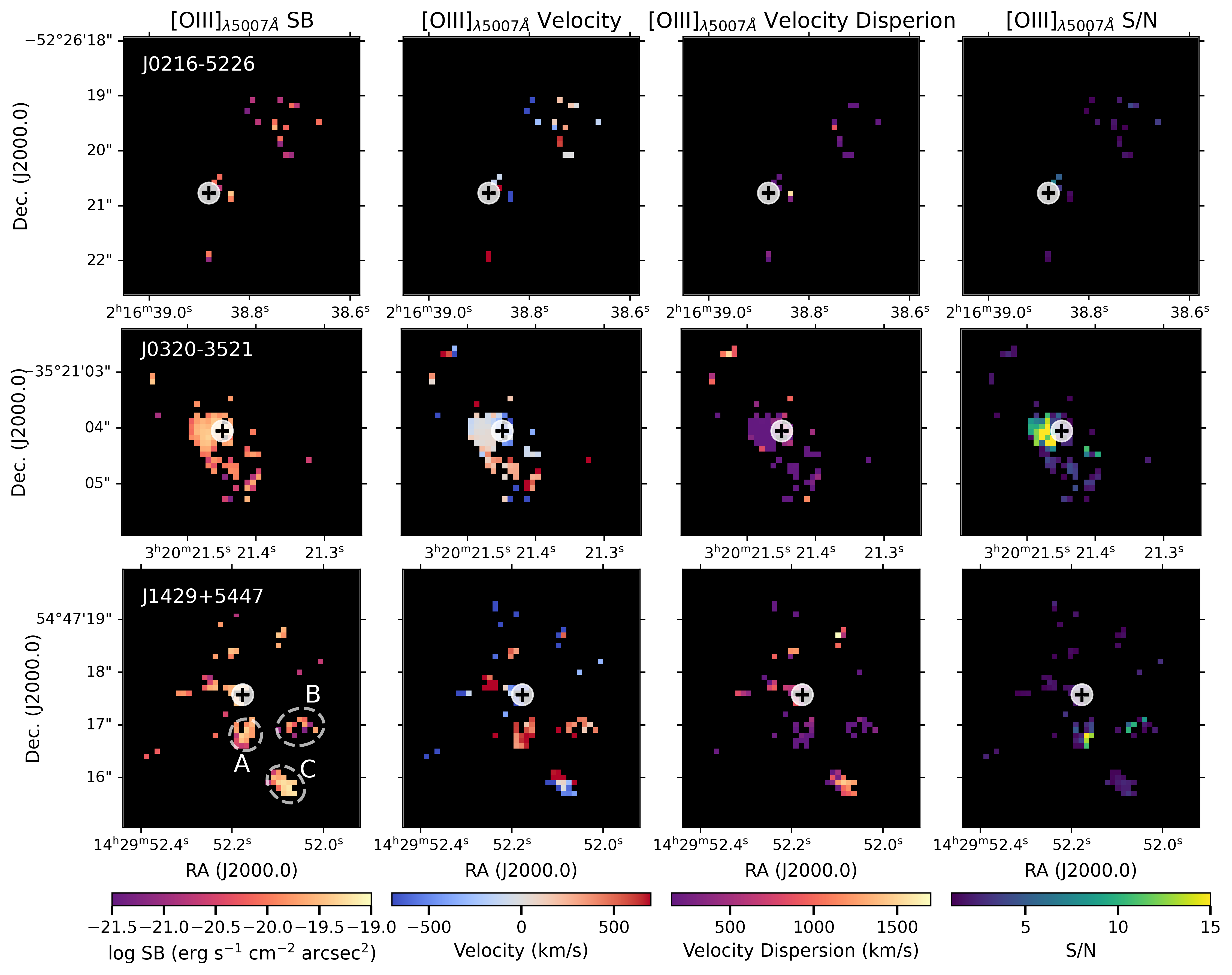}
}
\caption{\label{fig:OIII_maps} [\ion{O}{3}]$_{\lambda 5007}$ maps of J0216$-$5226 (top), J0320$-$3521 (middle), J1429+5447 (bottom), showing the integrated surface brightness (moment 0; first column), velocity offset (moment 1; second column), and velocity dispersion (moment 2; third column). The fourth column shows the signal-to-noise pixel map, with all maps showing only pixels with S/N$>1$ for presentation purposes. The quasar location is indicated by a black cross. }
\end{figure*}

The discovery of this strong emission line region close to J0320$-$3521 is especially interesting in combination with the indications of a high (possibly super-Eddington) accretion rate. If the emission-line region originates from an interacting companion galaxy, tidal interactions between the systems could be driving rapid gas inflow towards the nucleus (e.g., \citealt{Hernquist1989Natur.340..687H, Springel2005ApJ...620L..79S}), which can both trigger starbursts and feed black hole growth \citep{Sanders1988ApJ...325...74S, Mihos1992ApJ...400..153M, Mihos1994ApJ...425L..13M, DiMatteo2005Natur.433..604D}. Theoretical studies suggest that massive BH systems with frequent gas inflows can support long-lasting super-Eddington accretion (e.g., \citealt{Lupi2024A&A...686A.256L, Pezzulli2016MNRAS.458.3047P, Trinca2022MNRAS.511..616T, Trinca2024arXiv241214248T}), however, BH feedback processes can counteract this by efficiently suppressing gas accretion and regulating its growth \citep{DiMatteo2005Natur.433..604D, Springel2005ApJ...620L..79S, Hopkins2008ApJS..175..356H, Caleno2026arXiv260604081C}. 
From the current NIRSpec IFU data of J0320$-$3521, we cannot yet distinguish between an interacting companion galaxy and ionized gas associated with the quasar. Therefore, deeper data of its environment and host galaxy are necessary to understand the exact nature of this system.

\subsection{Complex environment of J1429+5447}

\begin{figure*}
\centering 
{\includegraphics[width=0.95\textwidth]{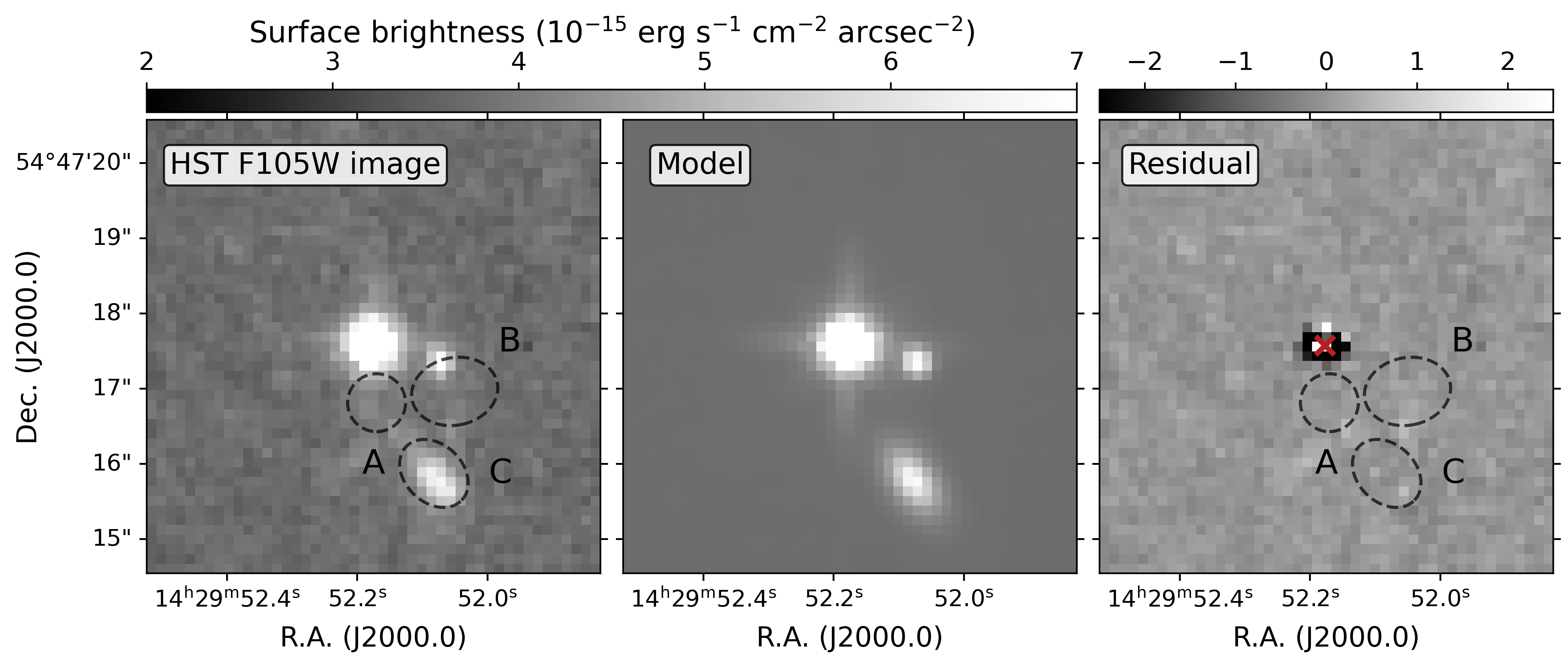}
} 
\caption{\label{fig:hst_im} HST F105W image of J1429+5447 confirming the presence of two neighboring sources in the projected plane. The locations of the three emission line regions found in the NIRSpec cube are indicated by the black ellipses labeled A--C. Middle: Model created from the HST image using \textsc{GALFIT}, including two point sources and one sersic source profile. Right: The residual of the model-subtracted HST image. The quasar location is indicated with a red cross.}
\end{figure*}

The resulting composite intensity map of J1429+5447 is shown in the right panel of Fig.~\ref{fig:rgb_maps}. We identify the presence of three distinct emission line regions in the IFU cube labeled as A, B, and C. 
The region around J1429+5447 has also been imaged using the Wide-Field Camera 3 (WFC3) on HST in F105W and F140W filters (see Sect.~\ref{subsection:hst_im}). The F105W image is shown in the left panel of Fig.~\ref{fig:hst_im}, and clearly shows the presence of an extended source at the location of component C and a point source close to component B. Both of these sources are also detected in the F140W image. We create a PSF model from unresolved stars in the image itself, and then make use of the \textsc{GALFIT} software \citep{Peng2002} to model the three sources in the image.
We define a flat background component with two point sources and a sersic profile for the extended source. The \textsc{GALFIT} model and model-subtracted HST image are also shown in Fig.~\ref{fig:hst_im}. 
While there is some marginal residual flux at the quasar location from the imperfect PSF subtraction, this model fits the data well. Therefore, we find no evidence for continuum emission in the HST image at the locations of A and B, while component C is clearly detected. We note that the point source west of J1429+5447 in the HST image (of $m_{F105W} \approx 24.7$ mag) is not detected in the collapsed NIRSpec IFU cube. The nature of this source remains unclear, although it is plausibly either a foreground star or a faint compact galaxy.

Previous observations of J1429+5447 with the Northern Extended Millimetre Array (NOEMA) and Expanded Very Large Array (EVLA) report the presence of two components of [\ion{C}{2}] emission \citep{Khusanova2022A&A...664A..39K} and CO(2-1) emission \citep{Wang2011ApJ...739L..34W}. \cite{Wang2011ApJ...739L..34W} finds two peaks spatially separated by 1.2\arcsec, which is speculated to indicate a possible major merging system. The location of this second CO(2-1) component (indicated in Fig.~\ref{fig:rgb_maps} by a pink cross) does not coincide with any obvious feature in our NIRSpec IFU data. The NOEMA data presented in \cite{Khusanova2022A&A...664A..39K} shows bright [\ion{C}{2}] emission at $z_{[\text{C}\textsc{ii}]}=6.190\pm0.004$, which is best fit by two Gaussian components, which may indicate a galaxy merger or AGN-driven outflows. However, due to the beam size of 1.6\arcsec$\times$1.4\arcsec, these were not spatially resolved, and therefore they cannot be compared directly to the emission-line regions in our IFU cube.

Finally, we optimally extract the spectra of the three emission line regions from the PSF-subtracted IFU cube at sky locations 14:29:52.17 +54:47:16.81 (region A), 14:29:52.05 +54:47:16.96 (region B), and 14:29:52.08 +54:47:15.87 (region C). Their spectra zoomed in on the wavelength regions of H$\beta$, [\ion{O}{3}], and H$\alpha$ at the quasar redshift are shown in Fig.~\ref{fig:IFU_companions_spectra}. Our combined findings and their interpretations are as follows:

\begin{itemize}
\setlength\itemsep{0.05em}
\item Component A, located at 0.77\arcsec\, (4.3 kpc projected distance) South of the quasar, shows H$\beta$, [\ion{O}{3}]$_{\lambda5007}$, and H$\alpha$ emission lines with a relative velocity shift of $\sim800$ km s$^{-1}$ with respect to the quasar. The [\ion{O}{3}] and H$\alpha$ line have a FWHM of 290$\pm$20 km s$^{-1}$ and 600$\pm$120 km s$^{-1}$, respectively. However, we note that the H$\alpha$ line is detected with only S/N$\sim5$. Again, we place an upper limit on the [\ion{N}{2}]/H$\alpha$ ratio of $<1.2$. Together with an [\ion{O}{3}]/H$\beta$ ratio of 5.9, this does not distinguish between excitation from star formation and AGN on the BPT diagram. Furthermore, there is no clear detection of \ion{He}{2}, \ion{O}{3}$_{\lambda4363}$, [\ion{O}{1}]$_{\lambda6300}$, and [\ion{S}{2}$_{\lambda\lambda6716,6731}$]. Since these emission lines are narrow, redshifted, and spatially disconnected from the quasar, we speculate that this is a companion galaxy of J1429+5447. 

\item Component B located at 1.3\arcsec\, (7.0 kpc projected distance) West of the quasar is faint, however, the extracted spectrum shows a detection of the H$\beta$ (S/N$\sim6$) and [\ion{O}{3}] (S/N$\sim9$) emission lines without any continuum. This emission could potentially be related to the point source detected in the HST F105W image, however, these are still offset by $\sim0.5$\arcsec ($\sim$2.8 kpc). The emission lines are narrow with a FWHM of 120$\pm$20 and 160$\pm$20 km s$^{-1}$ for H$\beta$ and [\ion{O}{3}], respectively. Again, there is no detection of \ion{He}{2}, [\ion{O}{1}]$_{\lambda6300}$, and [\ion{S}{2}$_{\lambda\lambda6716,6731}$]. We do note a detection of \ion{O}{3}$_{\lambda4363}$ with a FWHM of 121$\pm15$ km s$^{-1}$ and a redshift of $1933$ km s$^{-1}$.

\item Component C (at 1.9\arcsec\, South West of the quasar, $\sim10$ kpc projected distance) is continuum detected throughout the whole data cube without any clear emission line features. Together with its extended shape in the HST data, we interpret this as a foreground galaxy unrelated to the quasar. The best fit \textsc{GALFIT} model resulted in a magnitude of m$_{\text{F105W}} = 23.46$ mag. 
The detections of component C in Fig.~\ref{fig:rgb_maps} and \ref{fig:OIII_maps} are therefore continuum detections and do not reflect the specified emission lines. 
\end{itemize} 

Especially for components A and B, deeper JWST/NIRSpec observations are necessary to further investigate their physical origin.

\subsection{Faint emission-line region of J0216$-$5226}

The flux map of J0216$-$5226, including only pixels with S/N$>1$, shows only a single tentatively detected emission-line structure about $\sim1$\arcsec\, North-West from the quasar (see top panel in Fig.~\ref{fig:OIII_maps}). This structure virtually disappears when the S/N cut is increased to 2 and its extracted spectrum from the PSF subtracted IFU cube is dominated by noise without any clear emission lines. However, it does visually show up in a similar location in multiple integrated line maps, including H$\beta$, He\textsc{ii}$_{\lambda4686}$, [N\textsc{ii}]$_{\lambda6584}$, [\ion{O}{3}]$_{\lambda4363,5007}$, and [S\textsc{ii}]$_{\lambda6717,6731}$, suggesting its not an artifact. To confirm and investigate the nature of this emission region, deeper IFU observations are necessary. 

\section{Radio properties}
\label{sec:radio}

Two out of three quasars in our extreme $R_{\text{FeII}}$ outlier sample, J0320$-$3521 and J1429+5447, are well-known radio-loud quasars, while typically only 10\% of the quasar population is radio-loud. This raises the question of whether their radio-loudness is connected to their high accretion rates and metallicities. In general, radio-loud quasars occupy a similar parameter to radio-quiet quasars across a range of observational properties, such as their intrinsic UV to optical SED (e.g., \citealt{Gaskell2004ApJ...616..147G, Richards2006ApJS..166..470R}) and spectral properties (e.g., \citealt{Shen2011ApJS..194...45S}). While radio-loud quasars span a broad range of black hole masses and accretion rates, recent studies by \cite{Yue2025MNRAS.537..858Y} and \cite{Jackson2026MNRAS.546ag065J} have demonstrated that on a population level at low-$z$, powerful radio jets are preferentially associated with quasars with high black hole masses and/or high accretion rates. The rarity of radio-loud quasars at high-$z$ has prevented a similar statistical analysis at $z\gtrsim6$. 
The radio properties of both J0320$-$3521 (e.g., \citealt{Ighina2025ApJ...990L..56I}) and J1429+5447 (e.g., \citealt{Wang2011ApJ...739L..34W, Frey2011VLBI, Shao2020GMRT}) have been studied in previous works. Here we summarize those findings and remodel the radio spectrum of J1429+5447 with new ultra-low frequency data from the LOFAR LBA Sky Survey (LoLLS) at 54 MHz \citep{deGasperin2023arXiv230112724D}.

We note that J0216$-$5226 is not detected in either the RACS or the Evolutionary Map of the Universe survey (EMU; \citealt{Norris2011PASA...28..215N}). Therefore, we can only place a 3$\sigma$ limit on its radio luminosity of $L_{887\text{MHz}} \lesssim 1.9\times10^{26} $ W Hz$^{-1}$ from RACS and $L_{944\text{MHz}} \lesssim 2.3\times10^{25} $ W Hz$^{-1}$ from EMU, assuming a typical spectral index of $\alpha=-0.7$ (e.g., \citealt{Hardcastle2016MNRAS.462.1910H}). Due to its sky location, J0216$-$5226 is not covered by other large-scale radio surveys. If emitting in radio, its radio luminosity is more than $\sim25\times$ lower than J0320$-$3521 and J1429+5447. 
Furthermore, taking the 3$\sigma$ limit on the radio flux at 944 MHz and converting this to 5 GHz rest-frame, together with its previously measured M$_{4400A}$ value from SED fitting of $-25.96\pm0.13$ mag \citep{Gloudemans2021A&A...656A.137G}, gives an estimated 3$\sigma$ upper limit on its radio-loudness of $R \lesssim 4$. This upper limit suggests that J0216$-$5226 is radio-quiet. 
 
\subsection{Radio spectrum of J1429+5447}
\label{sec:radio_properties_J1429}

J1429+5447 is one of the radio-loudest quasars at $z>6$ with both its radio and submillimeter emission extensively studied (e.g., \citealt{Wang2011ApJ...739L..34W, Frey2011VLBI, Shao2020GMRT, Khusanova2022A&A...664A..39K, Li2024ApJ...962..119L}). High-resolution VLBI observations at GHz frequencies from \cite{Frey2011VLBI} have revealed the compact nature of the radio emission with a tentatively resolved structure of $<100$ pc towards the North-East. This projected jet direction does not align with any of the detected ionized regions in the NIRSpec IFU cube (see Sect.~\ref{sec:outflow_signatures}).  
Furthermore, its radio spectrum has been reported to be steep with $\alpha\sim - 0.6$ to $-0.8$ \citep{Frey2011VLBI, Shao2020GMRT}\footnote{We use the definition $S_{\nu} \propto \nu^{\alpha}$, with $\nu$ the frequency and $\alpha$ the spectral index.} and a suggested radio-loudness of $R \sim 100-200$ \citep{Banados2015ApJ...804..118B, Shao2020GMRT}. J1429+5447 has also been extensively studied at X-ray wavelengths, revealing a high X-ray luminosity ($L_{X} = 2.3^{+0.6}_{-0.5}\times10^{46}$ erg s$^{-1}$; \citealt{Migliori2023MNRAS.524.1087M}, see also \citealt{Medvedev2020MNRAS.497.1842M}) and significant variability \citep{ Marcotulli2025ApJ...979L...6M}.

Due to the high redshift of J1429+5447 of $z\approx6.16$, probing low radio frequencies is crucial to constrain the rest-frame GHz spectral shape. Therefore, we remodel its radio spectrum including new low-frequency detections from LoTSS-DR3 (144 MHz, \citealt{Shimwell2025LoTSSDR3}) and LoLLS-DR1 (54 MHz; \citealt{deGasperin2023arXiv230112724D}).
We combine these measurements with the multi-frequency radio flux densities reported in literature, including GMRT (323 MHz; \citealt{Shao2020GMRT}), VLA FIRST (1.4 GHz; \citealt{Becker1994ASPC...61..165B}), VLASS (2-4 GHz, \citealt{Lacy2020PASP..132c5001L}), EVN VLBI (1.6 and 5 GHz, \citealt{Frey2011VLBI}), EVLA (32 GHz; \citealt{Wang2011ApJ...739L..34W}). Specifically, the VLASS measurement of J1429+5447 was obtained from the Epoch 1 QuickLook catalog version 3.1\footnote{\url{https://cirada.ca/vlasscatalogueql0}}. Furthermore, J1429+5447 is not included in the LoLLS-DR1 catalog, however, clearly detected in the image. Therefore, we obtained its flux density using the Python Blob Detector and Source Finder (\textsc{PyBDSF}; \citealt{Mohan2015ascl.soft02007M}). The offset between the optical quasar and LoLLS radio detection is 4.3\arcsec, which is within the expected uncertainty given the surveys resolution of 15\arcsec\, and S/N of $\sim3$. For the other low-frequency measurement from the LoTSS survey (at 144 MHz), the measured offset is 0.23\arcsec\, with a confident S/N$\sim10$ detection. 
We note that these radio observations have been taken at wildly different resolutions ($\sim$mas to arcseconds). The highest resolution observation at 1.6 GHz from \cite{Frey2011VLBI} suggests the jet is compact with $<100$ pc, however, extended (lobe) emission is potentially missed at high frequencies and high resolution due to low surface brightness. To take into account any systematics, we add a 10\% flux error in quadrature to all reported flux density errors. All radio measurements used for SED fitting are summarized in Table~\ref{tab:radio_points}. 

\begin{deluxetable*}{lllll}
\tablewidth{0pt}
\tablecaption{Radio measurements of J1429+5447 used for SED fitting \label{tab:radio_points}}
\tablehead{
\colhead{Frequency} &
\colhead{Total flux density} &
\colhead{Telescope (Survey)} &
\colhead{Reference} 
\\[-4pt]
\colhead{(GHz)} &
\colhead{(mJy)} &
\colhead{} &
\colhead{}
}
\startdata
0.54 & $8.20\pm3.08$ & LOFAR (LoLLS-DR1) & \cite{deGasperin2023arXiv230112724D} \\
0.144 & $13.70\pm1.52$ & LOFAR (LoTSS-DR3) & \cite{Shimwell2025LoTSSDR3} \\ 
0.323 & $5.02\pm0.61$ & GMRT & \cite{Shao2020GMRT} \\
1.4 & $2.95\pm0.33$ & VLA (FIRST) & \cite{Becker1994ASPC...61..165B} \\
1.6 & $3.30\pm0.34^{(1)}$ & EVN VLBI & \cite{Frey2011VLBI} \\
2--4 & $1.82\pm0.27$ & VLA (VLASS) & \cite{Lacy2020PASP..132c5001L} \\ 
5.0   & $0.99\pm0.12$ & EVN VLBI & \cite{Frey2011VLBI}\\
32 & $0.26\pm0.03$ & EVLA & \cite{Wang2011ApJ...739L..34W} \\
\enddata
\tablecomments{A 10\% flux error has been added in quadrature to all reported flux density errors to account for systematics. (1) This is the sum of the flux densities in the VLBI components of $3.03\pm0.05$ mJy and $0.27\pm0.04$ mJy.}
\end{deluxetable*}

\begin{figure}
    \centering
    \includegraphics[width=\columnwidth]{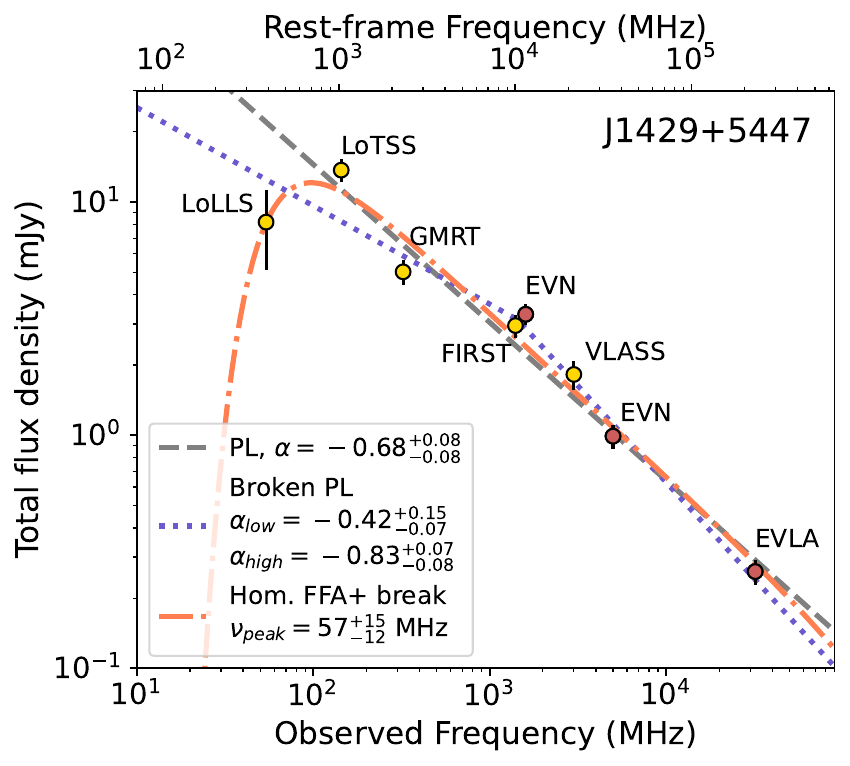}\vspace{-0.3cm}
    \caption{Radio spectral energy distribution of J1429+5447, ranging from 54 MHz to 32 GHz in the observed frame (bottom axis) and rest-frame (top axis). The red points highlight the flux density measurements from high-resolution radio images, while the yellow points are from low resolution images. The grey dashed line shows the best fitting power-law with a slope of $\alpha=-0.68\pm0.08$. However, the best fitting model is the broken power-law (purple dotted line), followed by the homogeneous Free-Free Absorption model with a break (orange line).} 
    \label{fig:J1429_radio_spec}
\end{figure}

The radio spectrum is shown in Fig.~\ref{fig:J1429_radio_spec}, which indicates a potential break or low-frequency turnover. Therefore, we model the radio spectrum using different types of models: a normal power-law (PL), broken PL, and absorption models. The two main types of absorption models that explain the turnover in radio spectra are Free-Free Absorption (FFA) and Synchrotron Self Absorption (SSA). We refer to \cite{callingham2015ApJ...809..168C} for details of these models. In short, the FFA model assumes that non-thermal radio emission is absorbed by an internal or external ionized screen, which is either homogeneous or inhomogeneous. In the SSA model, the same relativistic electrons that emit the synchrotron radiation also absorb photons at low frequency. The turnover frequency is the transition from the optically thick (below the turnover) to optically thin regime. 

For the fitting procedure, we use an invariant Markov Chain Monte Carlo (MCMC) approach with 5000 iterations. The errors on each parameter are given by the 16th and 84th percentiles. The different model fits are compared using the Bayesian Information Criterion (BIC; see eq.~\ref{eq:bic}). The gray line in Fig.~\ref{fig:J1429_radio_spec} shows the best fitting simple PL model with a spectral index of $\alpha=-0.68\pm0.08$, which translates to a k-corrected radio-luminosity of $L_{1.4\text{GHz}} = 5.5\times10^{26}$ W Hz$^{-1}$. In addition, we consider a broken PL model, homogeneous SSA model, internal FFA model, and homogeneous FFA model (with and without high-frequency break). The best fitting model to the radio SED of J1429+5447 is the broken PL with a BIC value of 29. This is followed by the homogeneous FFA model including a break, with a BIC value of 33. The broken PL (with $\alpha_{\text{low}} = -0.42^{+0.15}_{-0.07}$ and $\alpha_{\text{high}} = -0.83^{+0.07}_{-0.08}$) is therefore favored by our modeling as $|\Delta\text{BIC}| > 4$. While having the fewest model parameters, the simple PL used in previous works results in a higher BIC value of 49. Our three best fitting models are shown in Fig.~\ref{fig:J1429_radio_spec}. The break in the broken PL spectrum is likely caused by radiative losses (i.e. inverse Compton losses) from an ageing electron population, causing a steepening of the spectrum (e.g \citealt{Jaffe1973A&A....26..423J, Murgia2003PASA...20...19M}). The break frequency of J1429+5447 is estimated to be $\nu_b = 1.4\pm0.7$ GHz in observed frame ($10.3\pm4.8$ GHz in rest-frame), and is expected to move to lower frequencies as time passes without the injection of fresh particles. Similarly, a break has been observed in the radio spectrum of PSO J352.4034–15.3373 (at $z= 5.832\pm0.001$) at a high frequency of 28.76 GHz in observed frame (196.46 GHz rest-frame; see \citealt{Rojas-Ruiz2021ApJ...920..150R, Rojas-Ruiz2025ApJ...985...34R}) with an estimated jet spectral ageing from cooling of t$_{\text{spec}}\sim580$ yrs. Furthermore, the radio spectrum of the $z=6.1$ blazar PSO J030947.49+271757.31 shows a break at a few tens of GHz in the observed frame \citep{Spingola2020A&A...643L..12S, Gloudemans2023A&A...678A.128G}.

To summarize, the current radio data of J1429+5447 suggests its radio jet is compact ($<100$ pc at 1.6 GHz; \citealt{Frey2011VLBI}) with its radio spectrum showing a potential spectral turnover at low-frequency, which could indicate that this is a young gigahertz peaked-spectrum (GPS; e.g., \citealt{fanti1990A&A...231..333F, Odea1991ApJ...380...66O}) source. 
Inferring the orientation of the radio jet is challenging, as observational evidence both supports and disfavors a blazar interpretation (i.e., a jet viewed close to face-on). On the one hand, its high radio loudness, X-ray brightness, and rapid X-ray variability are consistent with blazar-like properties \citep{Marcotulli2025ApJ...979L...6M}. On the other hand, blazars typically exhibit flat radio spectra ($\alpha>-0.5$; e.g., \citealt{Coppejans2017MNRAS.467.2039C}), and \cite{Medvedev2020MNRAS.497.1842M} found no strong evidence for radio variability in this source, arguing against a blazar nature. Overall, the current data do not provide meaningful constraints on the viewing angle. 

\subsection{Radio and X-ray emission of J0320$-$3521}
\label{sec:radio_properties_J0320}

J0320$-$3521 was originally selected by combining RACS 888 MHz observations with the Dark Energy Survey \citep{Ighina2023MNRAS.519.2060I}. Its radio and X-ray properties are presented in \cite{Ighina2025ApJ...990L..56I}, with radio properties similar to other high-$z$ radio quasars with a spectral index of $\alpha=-0.72\pm0.02$ and radio luminosity of $L_{1.4\text{GHz}} = 5.8\times10^{26}$ W Hz$^{-1}$, which are both very similar to J1429+5447. Its known radio spectrum spans a frequency range of 100 MHz to 9 GHz with no indication of a break or turnover. The 2.3 GHz observation of LBA-VLBI shows a marginally resolved jet structure pointing North-West (see Fig. 5 of \citealt{Ighina2025ApJ...990L..56I}), which is not aligned with the ionized region found in the NIRSpec IFU cube in Sect.~\ref{sec:outflow_signatures}. However, this detection is tentative ($\sim3\sigma$) and likely dominated by noise. Therefore, follow-up observations are necessary to confirm any extended jet emission and alignment with the observed emission-line region.

Interestingly, \cite{Ighina2025ApJ...990L..56I} find from Chandra observations that J0320$-$3521 is among the brightest X-ray sources at $z\gtrsim6$, similar to J1429+5447, with $L_{2-10 \text{keV}} = 1.8^{+2.2}_{-0.9}\times10^{46}$ erg s$^{-1}$, and with an ultra-steep X-ray photon index of $\Gamma_X = 3.3\pm0.4$. Super-Eddington accretion is discussed as the most likely scenario explaining its steep X-ray spectrum and expected limit on the bolometric luminosity, consistent with multi-wavelength SED fitting and expectations from theoretical models (e.g., \citealt{Madau2024ApJ...976L..24M, Inayoshi2025PASJ...77..811I}). However, in the case of jetted quasars both the jet and corona contribute to the X-ray emission, complicating identifying the exact origin of the X-ray emission. We refer to \cite{Ighina2025ApJ...990L..56I} for more discussion.
We highlight that this work supports the potential super-Eddington nature of J0320$-$3521 by its location in the EV1 plane and the derived Eddington accretion ratios of $\lambda_{\text{Edd}} = 0.5-6.0$ (see Sect.~\ref{sec:spec_analysis}).

\section{Summary}
\label{sec:summary}

To summarize, we identify three high-$z$ quasars in the JWST \textsc{Aether} survey as outliers on the quasar main sequence. Their rest-frame optical spectra exhibit remarkably strong \ion{Fe}{2} features and weak [\ion{O}{3}] and H$\beta$, suggesting that these chemically enriched and potentially highly accreting quasars in the first Gyr of our Universe. Their measured rest-frame optical \ion{Fe}{2} strengths ($R_{\text{\ion{Fe}{2}},\lambda4570} \equiv \text{EW}_{\text{\ion{Fe}{2}}}/\text{EW}_{\text{H}\beta}$) of $\sim2.8-3.6$ are within the top 3\% of the SDSS quasar distribution at $0\lesssim z \lesssim 1$. However, we find that the \cite{Park2022ApJS..258...38P} \ion{Fe}{2} template provides a better fit to the spectral features at $<5600$ \AA, which resulted in $R_{\text{\ion{Fe}{2}},\lambda4570}$ values of $\sim2.5-2.7$. Analysis of previously taken ground-based spectroscopic data with Gemini/GNIRS (for J1429+5447) and Magellan/FIRE (for J0216$-$5226) supports this scenario of quick chemical enrichment, with high \ion{Fe}{2}/\ion{Mg}{2} ratios of $\sim6$.

The estimated H$\alpha$-based black hole masses are $\sim0.15-7\times10^9$ M$_{\odot}$, similar to the general high-$z$ quasar population, with derived Eddington ratios between $\sim0.09-0.39$ for J0216$-$5226, $\sim0.5-6.0$ for J0320$-$3521, and $\sim0.14-0.60$ for J1429+5447. These measurements suggest possible super-Eddington accretion in the case of J0320$-$3521, which is supported by its extreme X-ray properties as argued by \cite{Ighina2025ApJ...990L..56I}.
The BH mass estimates and Eddington ratios derived from H$\beta$ are similar with $\sim0.2-3\times10^9$ M$_{\odot}$. While H$\beta$ generally provides more reliable correlations, in these cases the H$\beta$ lines are weak and difficult to model together with the strong \ion{Fe}{2} pseudocontinuum features. Overall, these estimates of the BH masses and Eddington ratios depend heavily on the scaling relation, and \ion{Fe}{2} template used, and BH masses are known to be overestimated in \ion{Fe}{2}-strong quasars. These values therefore remain uncertain and should be used with caution. The location of these quasars on the fundamental plane suggests these three quasars are promising candidates for super-Eddington accreters.

The NIRSpec IFU data cubes furthermore allow for investigating the immediate quasar surroundings (within $\sim10$ kpc in projected distance). The integrated intensity maps of J1429+5447 indicate a complex environment with 2 possible companion or outflow features, and 1 foreground galaxy, as confirmed by HST WFC3 imaging. Also, the IFU data of J0320$-$3521 shows a distinct emission line region within $\sim$2 kpc of the quasar in projected distance, which could be a companion or ionized gas region.  
Finally, the IFU cube of J0216$-$5226 revealed a faint emission line region north-west of the quasar. However, due to the low S/N, the source causing the emission could not be identified. Deeper NIRSpec/IFU and potentially sub-mm observations are necessary to further examine the nature of these emission line regions.

Two of our quasars, J0320$-$3521 and J1429+5447, are radio-loud, while J0216$-$5226 remains undetected in current large-sky radio surveys, with upper limits suggesting J0216$-$5226 is radio-quiet ($R_{5\text{GHz}} \lesssim4$). The radio properties of both radio-loud quasars have previously been studied in literature, however, in this work, we remodel the radio spectrum of J1429+5447 including a recently obtained ultra-low frequency measurement at 54 MHz, which hints at a potential turnover in the radio spectrum. However, the modelling favors a broken power-law at an observed break frequency of 1.4 GHz ($\sim$10 GHz in rest-frame), caused by radiative losses from an ageing electron population. Comparing the ionized gas regions in our NIRSpec/IFU cubes to previous high-resolution radio observations, we do not find any evidence for radio jet alignment. 

This paper highlights three extreme outliers on the quasar main sequence already in place at $z>6$, as unveiled by recent JWST observations, which confirms quick chemical enrichment of quasars in the early Universe. Despite the small sample size, our results reveal marked diversity in environment and radio-jet properties among quasars that display similar optical spectral features. 
The exceptional wealth of data from the \textsc{Aether} survey will establish a new benchmark sample of quasar properties at $z\geq5.7$, and can provide necessary statistical evidence to investigate whether radio jets are preferentially generated by the most highly accreting quasars in the early Universe, and to disentangle the roles of black hole physics and environment. 

\begin{acknowledgments}

We thank the referee for a careful reading of the manuscript and for constructive comments that improved the paper.

E.P.F. is supported by the international Gemini Observatory, a program of NSF NOIRLab, which is managed by the Association of Universities for Research in Astronomy (AURA) under a cooperative agreement with the U.S. National Science Foundation, on behalf of the Gemini partnership of Argentina, Brazil, Canada, Chile, the Republic of Korea, and the United States of America. 

RD acknowledges support from the PRORIS 2025 program ``COSMOWebb'' and from
the INAF RSN1 minigrant 2024 ``The interstellar medium at high redshift'.

C.M. acknowledges support from Fondecyt Iniciación grant 11240336 and the ANID BASAL project FB210003. 

SO acknowledges support from the JWST programs JWST-GO-4056 and JWST-GO-05645 provided by NASA through grants from the Space Telescope Science Institute, which is operated by the Association of Universities for Research in Astronomy, Inc., under NASA contract NAS5-03127.

This work is based on observations made with the NASA/ESA/CSA James Webb Space Telescope. The data were obtained from the Mikulski Archive for Space Telescopes at the Space Telescope Science Institute, which is operated by the Association of Universities for Research in Astronomy, Inc., under NASA contract NAS 5-03127 for JWST. These observations are associated with the Cycle 3 JWST Survey \#5645.

Some of the data presented in this paper were obtained from the Mikulski Archive for Space Telescopes (MAST) at the Space Telescope Science Institute. The specific observations analyzed can be accessed via \dataset[https://doi.org/10.17909/zskj-ma68]{https://doi.org/10.17909/zskj-ma68}. STScI is operated by the Association of Universities for Research in Astronomy, Inc., under NASA contract NAS5–26555. Support to MAST for these data is provided by the NASA Office of Space Science via grant NAG5–7584 and by other grants and contracts.

Support for the JWST program JWST-GO-05645 was provided by NASA through grants from the Space Telescope Science Institute, which is operated by the Association of Universities for Research in Astronomy, Inc., under NASA contract NAS 5-03127.

This work was enabled by observations made from the Gemini North telescope. The scientific community is honored to have the opportunity to conduct astronomical research on Maunakea in Hawai‘i. We recognize and acknowledge the very significant cultural role and reverence of Maunakea to the Kanaka Maoli (Native Hawaiians) community.

\end{acknowledgments}

\appendix

\section{Comparison FeII templates of H$\beta$+\ion{O}{3} complex}
\label{appendix:Hb_FeII_template_comparison}

\begin{figure*}[ht]
    \centering
    \includegraphics[width=1.0\textwidth]{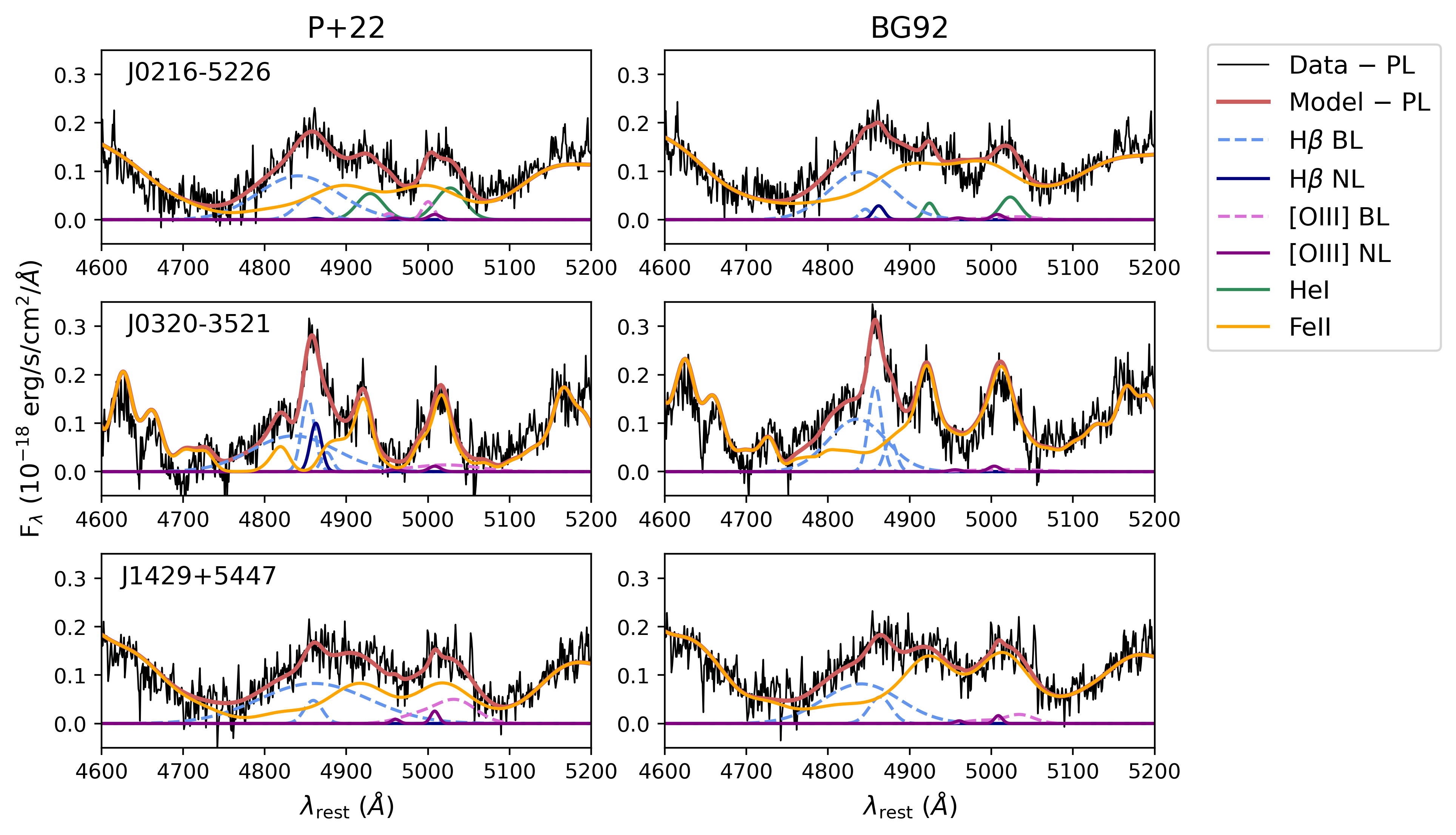}\vspace{-0.0cm}
    \caption{Zoom-in on the H$\beta$+\ion{O}{3} complex and the best-fitting models using the \cite{Park2022ApJS..258...38P} Mrk 493 template (P+22; left column) and the \cite{Boroson1992ApJS...80..109B} I Zw 1 template (BG92; right column). The best-fitting power-law continuum has been subtracted for clarity. The inferred \ion{Fe}{2} strength depends on the adopted template, with P+22 generally producing lower \ion{Fe}{2} equivalent widths and lower $R_{\text{\ion{Fe}{2}},\lambda4570}$ values.}
    \label{fig:Hbeta_comparison_iron_templates}
\end{figure*}

In Fig.~\ref{fig:Hbeta_comparison_iron_templates}, we compare the best-fitting \ion{Fe}{2} models obtained using the \cite{Park2022ApJS..258...38P} Mrk 493 template and the \cite{Boroson1992ApJS...80..109B} I Zw 1 template. The corresponding measurements are listed in Table~\ref{tab:sculptor_results}. We find that the \ion{Fe}{2} equivalent widths are generally lower when adopting the P+22 template, resulting in systematically lower values of $R_{\text{\ion{Fe}{2}},\lambda4570}$. 

Table \ref{tab:line_components} summarizes the FWHM values and central wavelengths of each individual Gaussian component fit of the H$\beta$, [\ion{O}{3}], and H$\alpha$ emission lines. For both J0216$-$5226 and J0320$-$3521, several narrow- and broad-line components converge to the imposed upper- and lower bounds of the FWHM allowed in the fitting routine. This reflects a degeneracy between the narrow and broad velocity components that we cannot break at the spectral resolution and S/N of our data. These components are generally low in amplitude and do not contribute significantly to the total line profile. 

\begin{deluxetable*}{lcccccc}
\tablewidth{0pt}
\tablecaption{Individual emission line components from spectral fitting \label{tab:line_components}}
\tabletypesize{\footnotesize}
\tablehead{
\colhead{Parameter}
& \multicolumn{2}{c}{J0216$-$5226}
& \multicolumn{2}{c}{J0320$-$3521}
& \multicolumn{2}{c}{J1429+5447}\\
\cline{2-3} \cline{4-5} \cline{6-7}
\colhead{}
& \colhead{P+22 \& BG92} & \colhead{BG92}
& \colhead{P+22 \& BG92} & \colhead{BG92}
& \colhead{P+22 \& BG92} & \colhead{BG92}
}
\startdata
\cutinhead{H$\beta$}
FWHM, Comp.~A (km~s$^{-1}$) & $7167^{+419}_{-384}$ & $5817^{+133}_{-129}$ & $7537^{+304}_{-284}$ & $5147^{+93}_{-103}$ & $9804^{+436}_{-415}$ & $6830^{+222}_{-237}$ \\
Center, Comp.~A (\AA) & $35790$ & $35790$ & $34628$ & $34628$ & $34792$ & $34651$ \\
FWHM, Comp.~B (km~s$^{-1}$) & $2507^{+345}_{-311}$ & $1000^{(1)}$ & $1000^{(1)}$ & $1081^{+467}_{-81}$ & $1599^{+161}_{-171}$ & $1963^{+156}_{-152}$ \\
Center, Comp.~B (\AA) & $35897$ & $35824$ & $34918$ & $34918$ & $34784$ & $34816$ \\
FWHM, Comp.~C (km~s$^{-1}$) & --- & --- & $1168^{+97}_{-137}$ & $1157^{+64}_{-86}$ & --- & --- \\
Center, Comp.~C (\AA) & --- & --- & $34753$ & $34785$ & --- & --- \\
FWHM, NL (km~s$^{-1}$) & $1000^{(1)}$ & $1000^{(1)}$ & $1000^{(1)}$ & $1000^{(1)}$ & $717^{+87}_{-65}$ & $687^{+41}_{-47}$ \\
Center, NL (\AA) & $35950$ & $35945$ & $34821$ & $34789$ & $34807$ & $34810$ \\
\cutinhead{[\ion{O}{3}]$\lambda\lambda4960,5008$}
FWHM, BL (km~s$^{-1}$) &  $1000^{(1)}$ & $3080^{+124}_{-140}$ & $5000^{(1)}$ & $5000^{(1)}$ & $3457^{+374}_{-323}$ & $2485^{+217}_{-178}$ \\
Center, BL (\AA) & $36612$, $36966$ & $36856$, $37212$ & $35670$, $36015$ & $35670$, $36015$ & $35690$, $36035$ & $35690$, $36035$ \\
FWHM, NL (km~s$^{-1}$) & $1000^{(1)}$ & $1000^{(1)}$ & $1000^{(1)}$ & $1000^{(1)}$ & $717^{+87}_{-65}$ & $687^{+41}_{-47}$ \\
Center, NL (\AA) & $36672$, $37026$ & $36666$, $37021$ & $35520$, $35863$ & $35487$, $35830$ & $35506$, $35849$ & $35508$, $35852$ \\
\cutinhead{H$\alpha$}
FWHM, Comp.~A (km~s$^{-1}$) & $12687^{+485}_{-392}$ & $13561^{+542}_{-453}$ & $8517\pm131$ & $9257^{+156}_{-143}$ & $12764^{+589}_{-506}$ & $14346^{+606}_{-585}$ \\
Center, Comp.~A (\AA) & $48426$ & $48420$ & $46748$ & $46748$ & $46827$ & $46809$ \\
FWHM, Comp.~B (km~s$^{-1}$) & $4268^{+144}_{-132}$ & $4372^{+140}_{-137}$ & $1485^{+74}_{-64}$ & $1985^{+83}_{-82}$ & $4042^{+157}_{-156}$ & $4091^{+121}_{-119}$ \\
Center, Comp.~B (\AA) & $48405$ & $48405$ & $46748$ & $46999$ & $46985$ & $46981$ \\
FWHM, Comp.~C (km~s$^{-1}$) & --- & --- & $1082^{+77}_{-82}$ & $1767^{+121}_{-101}$ & --- & --- \\
Center, Comp.~C (\AA) & --- & --- & $46924$ & $46748$ & --- & --- \\
FWHM, NL (km~s$^{-1}$) & $1000^{(1)}$ & $1000^{(1)}$ & $1000^{(1)}$ & $1000^{(1)}$ & $717^{+87}_{-65}$ & $687^{+41}_{-47}$ \\
Center, NL (\AA) & $48533$ & $48526$ & $47008$ & $46965$ & $46990$ & $46993$ \\
\enddata
\tablecomments{NL = narrow line; BL = broad line; Comp. A/B/C denote the individual broad-line profile components. A value of `---' indicates that a component was not included in the fit (e.g., Comp.~C for the two-Gaussian broad-line fits). (1) For these components, the fit converges to the upper or lower limit allowed for the FWHM, which reflects a degeneracy between the narrow and broad velocity components that we cannot break at the spectral resolution and S/N of our data. The contribution of these components to the overall total line profile is generally small as demonstrated in Fig.~\ref{fig:Hbeta_comparison_iron_templates}.}
\end{deluxetable*}

\onecolumngrid

\section{Moment maps of H$\beta$ and H$\alpha$}

\begin{figure*}
\centering 
{\includegraphics[width=\textwidth]{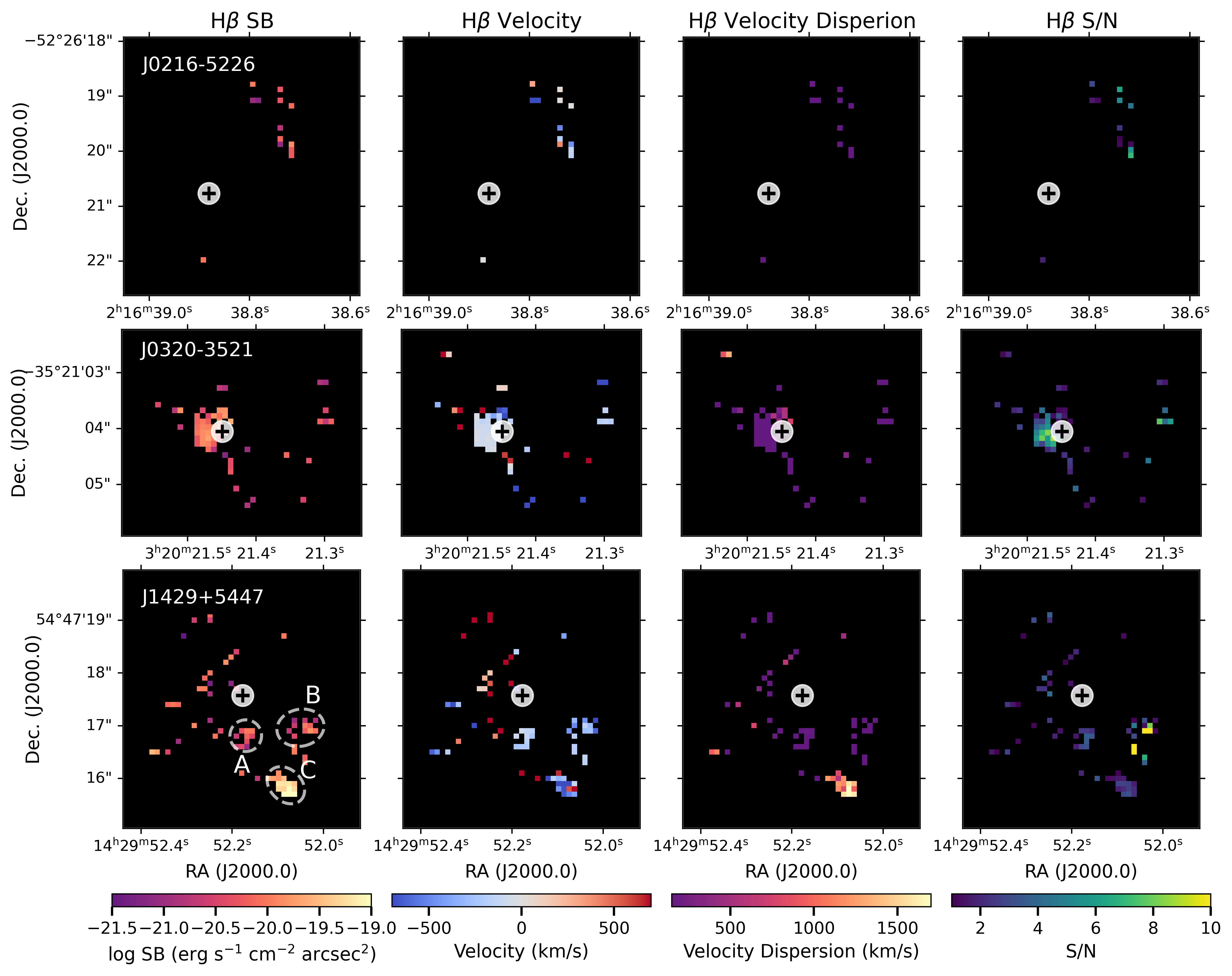}}
\caption{\label{fig:Hb_maps} H$\beta$ maps of J0216$-$5226 (top), J0320$-$3521 (middle), J1429+5447 (bottom), showing the integrated surface brightness (moment 0; first column), velocity offset (moment 1; second column), velocity dispersion (moment 2; third column), and signal-to-noise pixel map (fourth column). The quasar location is indicated by a black cross. The maps only show pixels with S/N$>1$.}
\end{figure*}

\begin{figure*}
\centering 
{\includegraphics[width=\textwidth]{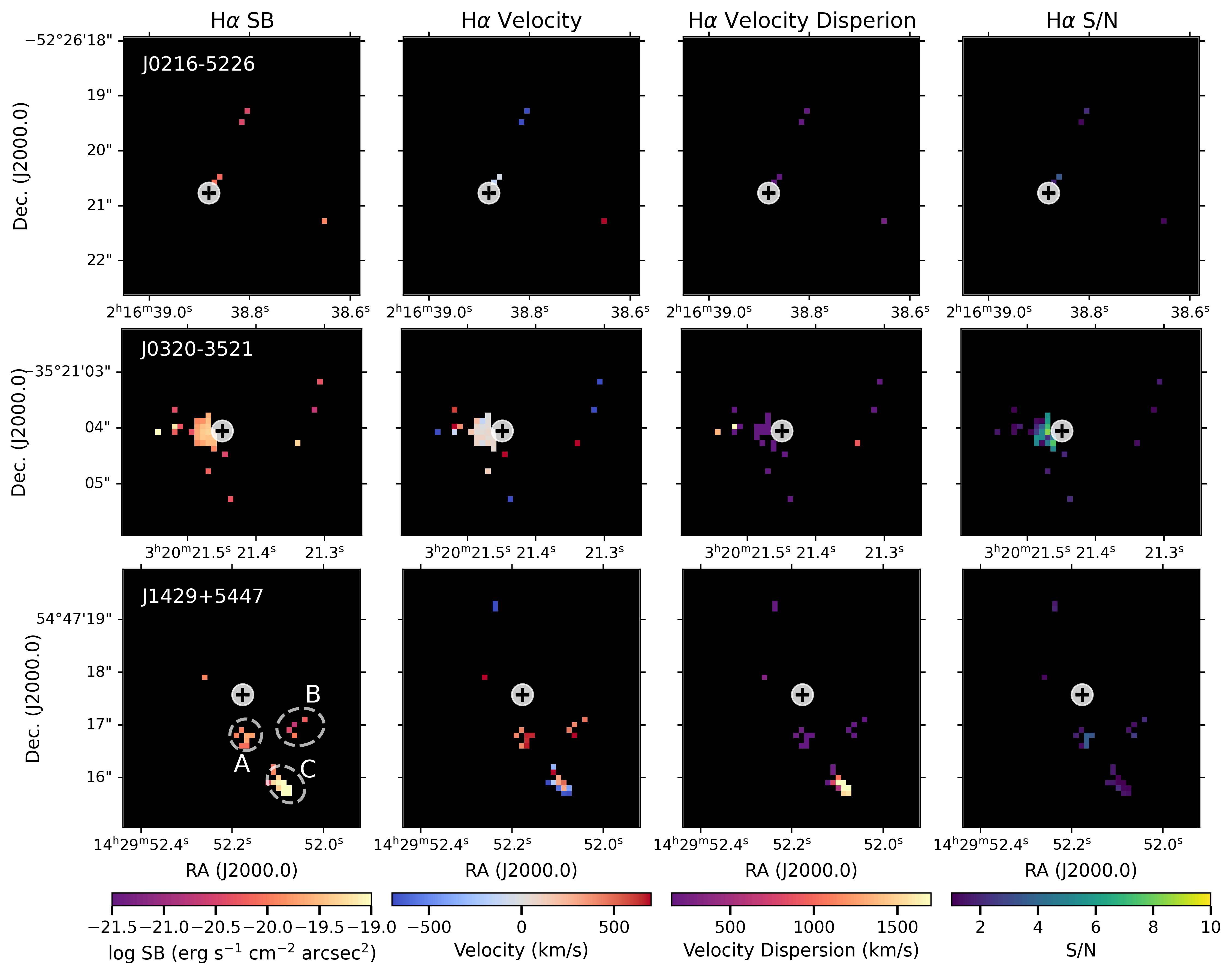}}
\caption{\label{fig:Ha_maps} Same as Fig.~\ref{fig:Hb_maps} for the H$\alpha$ emission line.}
\end{figure*}

Figure \ref{fig:Hb_maps} and \ref{fig:Ha_maps} show the H$\beta$ and H$\alpha$ moment maps, including surface brightness, velocity, and velocity dispersion, of the three quasars studied in this work. Pixels below S/N$<1$ are masked. The emission line regions identified for J1429+5447 in Sect.~\ref{sec:outflow_signatures} are highlighted in the bottom panel. As apparent from the maps, the tentatively detected region northwest of J0216$-$5226 (top panel) is also visible in H$\beta$, however, no coherent structure is observed in H$\alpha$. The emission line regions identified in the region of J0320$-$3521 and J1429+5447 are apparent in both maps. We interpret component C as a foreground galaxy. The moment maps at its location therefore do not trace genuine H$\beta$ and H$\alpha$ emission line kinematics, but instead reflect its underlying continuum emission.

\bibliography{main}{}
\bibliographystyle{aasjournalv7}

\end{document}